\RequirePackage{fix-cm}
\documentclass[twocolumn]{svjour3}          
\smartqed  

\usepackage{graphicx}
\usepackage{subcaption}
\usepackage{graphicx,color}
\usepackage{booktabs}
\usepackage{listings}

\usepackage{enumitem}
\newcommand{\subscript}[2]{$#1  #2$}

\lstdefinelanguage{SQL}{%
  alsoletter={-_:()\$\\},
  morekeywords={
   select,
   type,
   dataset,
   let,
   create,
   drop,
   dataverse,
   into,
   value,
   index,
   if,
   exists,
   use,
   closed,
   open,
   upsert,
   primary,
   key,
   autogenerated,
    from,
    where,
    and,
    or,
    in,
    as,
    group,
    by,
    join,
    on,
    with,
    update,
    set,
    delete,
    insert
  },
  basicstyle=\sf,
  keywordstyle=\textbf,
  identifierstyle=\texttt,
  commentstyle=\textit,
  literate={<=}{{\litleq}}1 {>=}{{\litgeq}}1 {//}{{\litdoubleslash}}1 
}[keywords,comments,strings]

\lstnewenvironment{sql}
   {\lstset{language=SQL,columns=flexible,basicstyle=\footnotesize,basewidth={.3em,.2em}}}
   {}

\begin{document}

\title{BAD to the Bone}
\subtitle{\underline{B}ig \underline{A}ctive \underline{D}ata at its Core}


\author{Steven Jacobs \and  
Xikui Wang \and 
Michael J. Carey \and 
Vassilis J. Tsotras \and 
Md Yusuf Sarwar Uddin
}


\institute{Steven Jacobs \at
              Univ. of California, Riverside \\
              \email{sjaco002@ucr.edu}       
           \and
Xikui Wang \at
              Univ. of California, Irvine \\
              \email{xikuiw@uci.edu}  
                        \and
Michael J. Carey \at
              Univ. of California, Irvine \\
              \email{mjcarey@ics.uci.edu} 
                         \and
Vassilis J. Tsotras \at
              Univ. of California, Riverside \\
              \email{tsotras@cs.ucr.edu} 
                         \and
Md Yusuf Sarwar Uddin \at
              Univ. of Missouri-Kansas City \\
              \email{muddin@umkc.edu} 
}

\date{Received:  23 July 2018 / Accepted: 27 April 2020}

\maketitle

\begin{abstract}
Virtually all of today's Big Data systems are \textit{passive} in nature, responding to queries posted by their users. 
Instead, we are working to shift Big Data platforms from passive to \textit{active}. In our view, a Big Active Data (BAD) system should continuously and reliably capture Big Data while enabling timely and automatic delivery of relevant information to a large pool of interested users, as well as supporting retrospective analyses of historical information. 
While various scalable streaming query engines have been created, their active behavior is limited to a (relatively) small window of the incoming data.

To this end we have created a BAD platform that combines ideas and capabilities from both Big Data and Active Data (e.g., Publish/Subscribe, Streaming Engines). It supports complex subscriptions that consider not only newly arrived items but also their relationships to past, stored data. Further, it can provide actionable notifications by enriching the subscription results with other useful data. Our platform extends an existing open-source Big Data Management System, Apache AsterixDB, with an \textit{active toolkit}. The toolkit contains features to rapidly ingest semistructured data, share execution pipelines among users, manage scaled user data subscriptions, and actively monitor the state of the data to produce individualized information for each user.

This paper describes the features and design of our current BAD data platform and demonstrates its ability to scale without sacrificing query capabilities or result individualization. 
\keywords{Big Data \and Big Active Data \and Active Data }
\end{abstract}

\section{Introduction} \label{introduction}

Work on Big Data management platforms has led to map-reduce based frameworks that provide after-the-fact Big Data analytics systems \cite{hadoop:website,pig:website,HivePaper,Spark} as well as NoSQL stores \cite{hbase,couchbase,mongodb} that focus primarily on scalable key-based record storage and fast retrieval for schema-less data. There are also modern platforms that seek to provide the benefits of both analytics and NoSQL \cite{alexandrov2014stratosphere,alsubaiee2014asterixdb}. While these systems generally scale well, they remain mostly ``passive'' in nature, replying with answers when a user poses a query.

With the ever-increasing amounts of data being generated daily by social, mobile, and web applications, as well as the prevalence of the Internet of Things, it is critical to shift from passive to ``active'' Big Data, overcoming barriers in ingesting and analyzing this sea of data and continuously delivering real time personalized information to millions of users. 
Past work on active databases \cite{dayal1988hipac,TriggerMan,hanson1996ariel,stonebraker1986design,ilprints7} was never built to scale to modern data sizes and arrival rates. 
Recently, various systems have been built to actively analyze and distribute incoming data; these include Publish/Subscribe systems \cite{EFG+03,Hojjat-middleware,MZV07,MBT07,saigaonkar2011publish,dynatops} and Streaming Query engines \cite{flink1,NiagaraCQ,chintapalli2016benchmarking,goldberg1992using,lee2016road}. However, these approaches have achieved a compromise by accepting functional constraints in order to scale (e.g, limiting queries to a recent window of data, or supporting a specific class of queries).

In contrast, we advocate for Big Active Data, an approach that aims to leverage modern Big Data Management in order to scale without giving up on data or query capabilities, and in particular to address the following \textit{BAD desiderata}:

\begin{enumerate}[label=(\subscript{D}{{\arabic*}})]
  \item Incoming data items might not be important in isolation, but in their \textbf{relationships} to other items in the data as a whole. Subscriptions need to consider \textbf{data in context}, not just newly arrived items' content.
  \item Important information for users is likely to be missing in the incoming data items, yet it may exist elsewhere in the data as a whole. The results delivered to users must be able to be \textbf{enriched} with other existing data in order to provide \textbf{actionable notifications} that are individualized per user.
   \item In addition to on-the-fly processing, later queries and analyses over the collected data may yield important insights. Thus, \textbf{retrospective Big Data analytics} must also be supported.
\end{enumerate}

To build a Big Active Data (BAD) platform,
we have leveraged the benefits of Apache AsterixDB, a modern Big Data Management System (BDMS) \cite{asterix:website,alsubaiee2014asterixdb} (namely, its scalability, declarative query language, flexible data model, and support for parallel data analytics), as well as borrowing key ideas underlying the active capabilities offered by existing Pub/Sub and streaming query systems.

A complete BAD platform will fully utilize and integrate all layers of the platform with all three goals in mind. In contrast, some related systems have been proposed or implemented (e.g., \cite{structured}) which seek to ``glue'' together separate existing platforms that accomplish different parts of these goals. Such systems (as seen through experimentation, e.g., \cite{grover2015data}) will suffer the disadvantages of (1) introducing complexity of communication and processing between systems that aren't built specifically to work in unison and (2) missing on potential performance gains by treating individual components (e.g., a permanent storage layer) as black boxes to communicate with. Creating a BAD platform presents the challenge of implementing the platform completely, but gives great gains in utilizing all layers with knowledge of the overall BAD vision.

The rest of the paper is organized as follows: Section \ref{overview} gives a high level overview of the objectives, capabilities, and needs of a BAD platform, and it introduces an example BAD application that we will examine for the bulk of the paper. In Section \ref{related} we discuss related work and the shortcomings of existing systems, which fall short in either capabilities or performance (or both) with respect to our BAD platform requirements. Since we aim to build a BAD system starting from an existing passive BDMS (namely AsterixDB), Section \ref{asterix} details the advantages offered by AsterixDB but also its shortcomings with respect to our active requirements.
In Section \ref{active-asterix} we introduce our Active Toolkit, which can be used to create a BAD platform. Section \ref{threelayers} delves into the three communication layers of BAD, used to maintain subscriber interests, discover complex data states of interest, and distribute this information to subscribers efficiently.
In Section \ref{glue} 
we briefly step aside to consider how one might build a BAD application today, in the absence of BAD, by combining multiple existing systems.
In Section \ref{experiments} 
we take a first look at BAD's performance characteristics using a synthetic workload inspired by our example application.
We also briefly look at how the same application might be built using existing passive Big Data technology (e.g., triggers) and the shortcomings of such an implementation.
Finally, Section \ref{conclusion} concludes the paper and describes our future plans.

\section{A BAD overview} \label{overview}

\begin{figure}[!ht]
  \centering
   \includegraphics[width=0.45\textwidth]{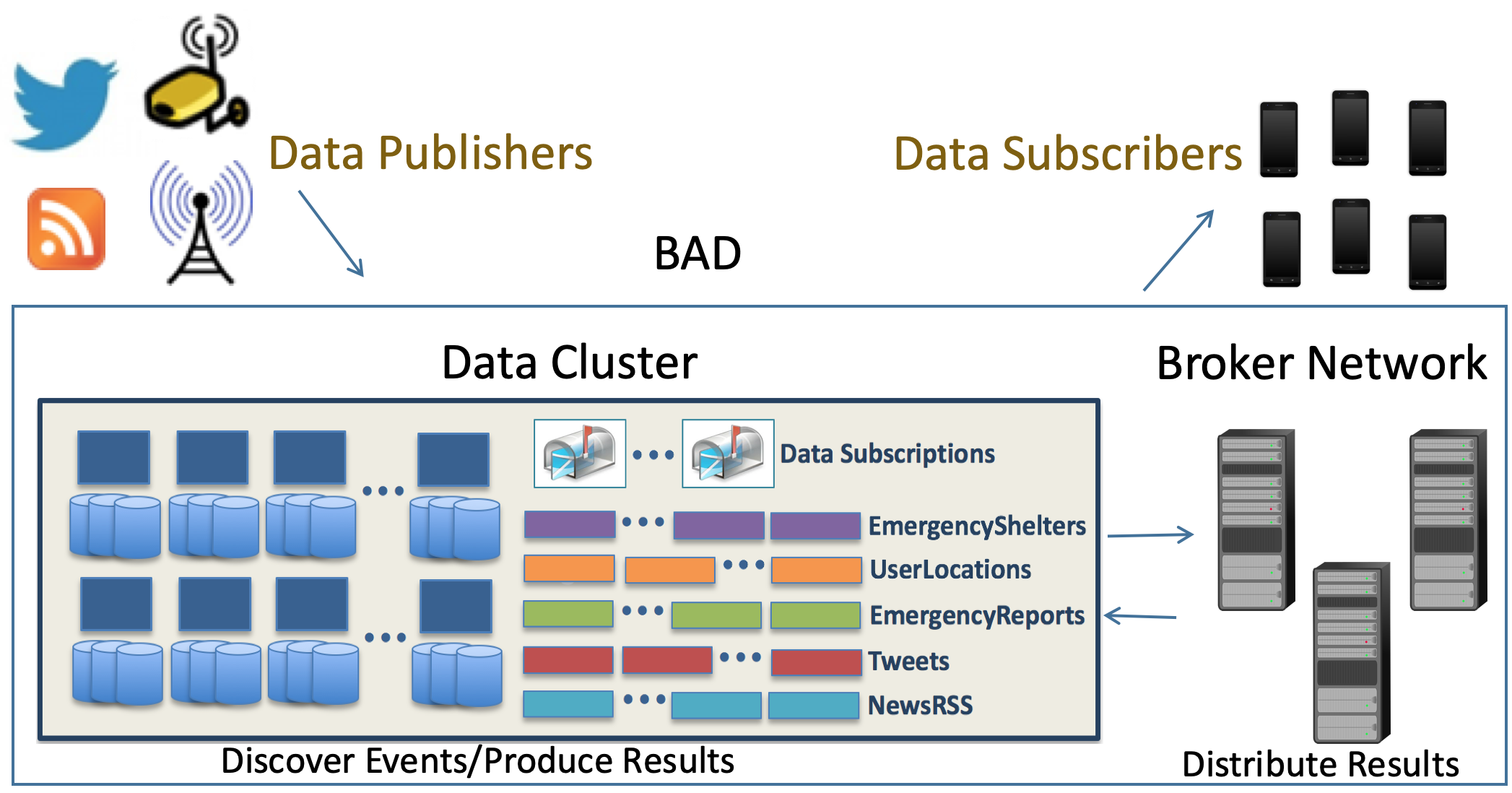}
    \caption{Big Active Data -- System Overview.}
  \label{sys}
\end{figure}

Figure \ref{sys} provides a 10,000 foot overview of our BAD Platform.
Outside of the platform, and the reason for its existence, are its data sources (Data Publishers) and its end users (Data Subscribers).
Within the system itself, its components provide two broad areas of functionality -- Big Data monitoring and management (handled by the BAD Data Cluster) and user notification management and distribution (handled by the BAD Broker Network).

\subsection{A BAD Application}
\label{sec:bad_application}
Consider as an example existing emergency notification services. 
One of the limitations of systems such as the USGS ShakeCast \cite{shakecast14} system is that their notifications are blanket statements for everyone (e.g., everyone receives the same flood warning or Amber Alert message). There is no individualization of messages to meet the needs of specific users (e.g., adding information relevant to the users' specific locations or needs). Such systems belong to the Pub/Sub system category, where users just get messages for topics of interest. 

In contrast, suppose a BAD system existed with the three capabilities sketched in Section \ref{introduction}. Rather than a user simply subscribing to emergency publications, now a user could ask something like ``when there is an emergency near my child, and it is a flash flood, please notify me, and provide me with the contact information for the security on-duty at the school, as well as nearby safety shelter locations.'' Suddenly a user is not getting a simple publication, but a rich set of data specifically relevant to the user, including the enrichment of the emergency information with security personnel schedules, local shelter information, etc.

We will use a similar example for demonstration and evaluation purposes for the bulk of this paper. Specifically, we will build a hypothetical example application that uses three data sources: \emph{UserLocations} (with records indicating the current location of each user), \emph{Reports} (containing a record for each emergency generated as emergencies occur), and \emph{Shelters} (holding the known locations of emergency shelters). 

\emph{UserLocations} and \emph{Reports} will be continuously ingested into the data cluster, 
whereas \emph{Shelters} will be loaded once and can be thought of as mainly a static, reference dataset (i.e., infrequently updated). We will focus on users who want to know about emergencies occurring near them in space and time, and providing those users with individualized shelter information based on their locations. Note how the three user needs described in Section \ref{introduction} apply to this example. The emergencies are only important to users if they are near the known reported location of the user, and the provided notifications are enriched with shelter information before delivery. The emergency reports data will continue to grow over time and can be analyzed later to gain historical insights to help with long-term emergency service planning.

It should be noted that this example is intended to serve as a simple (toy) example for illustrative purposes. Real potential use cases are varied and many. In addition to uses in emergency management, a few examples of possible applications include: (i) public safety, where one could monitor social media for various forms of concerning chatter near (or about) sensitive or public areas, or by certain “watched” individuals, to try and preempt mass shootings or other acts of terror; (ii) public health, where one could monitor social media comments and hospital reports to the CDC for patterns that could provide early warnings of infectious disease outbreaks; and (iii) business, where one might monitor a combination of customer service call records, other customer data, social media activity, and related product data for events that might forewarn a company about a potential impending departure of a valued customer.

\subsection{BAD Platform Prerequisites} \label{requirements}
A fully BAD Platform should take advantage of technologies and techniques that exist today for both Big Data performance and Scalable Delivery of results.

\textbf{Big Data performance}: 
For functionality and scalability, BAD should utilize the full capabilities of a modern BDMS; specifically, such systems can offer:

\begin{enumerate}
    \item A rich, declarative query language.
    \item Rich data type support that includes numeric, textual, temporal, spatial and semi-structured data.
    \item Capabilities for fast data ingestion.
    \item A data-partition-aware query optimizer.
    \item A dataflow execution engine for partitioned-parallel execution of query plans.
\end{enumerate}

Item (1) above, prevents BAD applications from being limited in query capability, while (2) and (3) allow for the variety and velocity of modern Big Data, as well as active spatial-temporal queries. (4) and (5) enable the scaling of the data volume and optimizing of active tasks to run in parallel across the BAD platform.

\textbf{Scalable Delivery of results}: Since a BAD data cluster is useless unless it can rapidly process data and its results actually reach its end users, a BAD platform must also offer the full capabilities of a Publish/Subscribe distributed network, including:

\begin{enumerate}
    \item Geo-distributed brokers that can scale dynamically to the demand of subscribers.
    \item Dynamic heuristics for handling large influxes of subscribers and results.
    \item Caching mechanisms designed with subscriber connectivity issues and commonality of interests in mind.
    \item Enhancements for communicating with the underlying BAD data cluster efficiently.
    \item Support for rapid ingestion of incoming events from various data sources.
    \item Low-latency requirements for delivery of results.
\end{enumerate}

\section{Related Work} \label{related}
Our work draws on work and ideas from modern Big Data platforms, early active database systems, and modern active platforms including both Publish/Subscribe systems and Streaming Query systems.

\subsection{Big Data Platforms}
First-generation Big Data management efforts resulted in various MapReduce-based \cite{mapreduce} frameworks and languages, often layered on Hadoop \cite{hadoop:website} for long-running data analytics; in key-value storage management technologies \cite{atikoglu2012workload,borkar2016have,conf/sosp/DeCandiaHJKLPSVV07,hyperdex} that provide simple but high-performance record management and access; and, in various specialized systems tailored to tasks such as scalable graph analysis \cite{dalvi2008keyword,low2012distributed,Malewicz:2010:PSL:1807167.1807184,yan2016big}
or data stream analytics \cite{abadi2003aurora,babu2001continuous,Dindar:2009:DDP:1559845.1559971,Gedik:2008:SSS:1376616.1376729,4812502,conf/sigmod/MarkowetzYP07}. With the exception of data streams (which limit query capabilities in order to scale), these developments have been ``passive'' in nature -- meaning that query processing, updates, and/or data analysis tasks have been scaled up to handle very large volumes of data, but these tasks run only when explicitly requested.

Several recent Big Data projects, including Apache Flink (Stratosphere) \cite{flink1,carbone2015apache}, Apache Spark \cite{website:spark,zaharia2016apache}, and Apache AsterixDB \cite{asterix:website,alsubaiee2014asterixdb}, have made strides in moving away from the tradition of the MapReduce paradigm, moving instead towards new approaches based on algebraic runtime engines. Nevertheless, these approaches maintain a mostly-passive approach. \emph{Data feed} mechanisms, such as those offered in AsterixDB \cite{grover2015data,DBLP:journals/pvldb/WangC19}, provide a step in the direction of becoming active, and we have advanced and evolved them to become part of our Active Toolkit.

Recent work using the lambda architecture \cite{lambda} design pattern seeks to provide a Big Data back-end as well as massive scale batch processing by combining a storage solution with a large-scale data query processing engine (typically triggered via batch jobs) in order to continuously ingest and analyze data. Though such solutions may fulfill some of the requirements for BAD, they do so by ``gluing'' several systems together, and they also focus on batch-queries for the sake of overall analytics, rather than on producing and delivering individualized results to scalable numbers of users.

\subsection{Active Data}
The key foundations for active data (ECA rules, triggers) were arguably laid by the HiPac Project \cite{dayal1988hipac}. Many other systems contributed to the work on ECA rules, including TriggerMan \cite{TriggerMan}, Ariel 
\cite{hanson1996ariel}, Postgres \cite{stonebraker1986design}, and Starburst \cite{ilprints7}. 
There are, however, two issues when directly applying past active techniques on Big Data.
First, triggers and ECA rules can be seen as a ``procedural sledgehammer'' for a system: when event A happens, perform action B. We seek a more declarative (optimizable) way of making Big Data active and detecting complex events of interest. Second, to the best of our knowledge, no one has scaled an implementation of triggers or ECA rules to the degree required for Big Data (in terms of either the number of rules or the scaled out nature of the data itself).

Work on Materialized View Maintenance (e.g., \cite{agrawal2009asynchronous,chirkova2012materialized,nikolic2014linview,quass1997line})
is also related to Active Data. 
Nevertheless, materialized view implementations have generally been designed to scale on the order of the 
the number of tables. 
Being more of a database performance tuning tool, the solutions developed in this area have not tried to address the level of scale that we anticipate for the number of simultaneous data subscriptions that should be the target for a BAD platform.

\subsection{Publish/Subscribe Systems}
In Pub/Sub Systems the data arrives in the form of publications, and these publications are of interest to specific users. 
Pub/Sub systems seek to optimize the problems of identifying the relevant publications and of delivering those publications to users in a scalable way. Early Pub/Sub systems were mostly topic-based (e.g., a user might be interested in sports or entertainment as publication topics). Modern Pub/Sub systems \cite{EFG+03,Hojjat-middleware,DBLP:conf/ipps/LiYKCL11,MBT07,saigaonkar2011publish} provide a richer, content-based subscription language, with predicates over the content of each incoming publication. Our BAD platform goes beyond this functionality in two ways:
First, whether or not newly arrived data is of interest to a user can be based on not only its content, but on its relationships to \emph{other} data. Second, the resulting notification(s) can include information drawn from other data as well. 

There has been some work done to enable Pub/Sub systems to cache data in order to provide a richer subscription language and result enrichment \cite{jin2003relational,qader2017dualdb,wang2017top}, but this research has largely relied on limiting the size of the cached data (e.g., by storing a window of recent history). This limitation prevents subscriptions from being applied to Big Data as a whole.

\subsection{Continuous Query Engines}
The seminal work on Tapestry \cite{goldberg1992using} first introduced Continuous Queries over append-only databases, including a definition of monotonic queries.
Most subsequent research has focused on queries over streaming data (e.g., STREAM \cite{DBLP:journals/debu/ArasuBBDIMNSTVW03}, Borealis \cite{Abadi05thedesign}, Aurora \cite{abadi2003aurora}, TelegraphCQ \cite{DBLP:conf/cidr/ChandrasekaranDFHHKMRRS03}, and PIPES \cite{kramer2004pipes}). These systems are typically implemented internally by building specialized data flows (``boxes and arrows'') to process query results as new data streams transiently through the system.
Related to both Pub/Sub and streaming data, the Distributed Reactive Programming model \cite{reactive} which some systems are starting to adopt seeks to create event-driven applications that coordinate and react to multiple events in order to produce state or event outputs.

Recently, more advanced algebraic streaming query engines (e.g., Storm, Flink, and Spark Streaming) \cite{chintapalli2016benchmarking} have been produced, which provide robust Big Data scale processing of incoming data, but they are still designed to work on incoming flows or windows, not providing a means of permanent Big Data storage. Structured Streaming on Spark \cite{structured} provides improvements for such systems, introducing a more user-friendly declarative API as well as providing the ability to join streams with static data sources. However, even at their best, streaming query systems are not designed to offer a subscription model for delivering individual results to a scalable number of users, instead outputting the results of a large job to a single log or database. As a result, they don't provide individualization or enrichment of results or a scaled user-based online delivery mechanism. Essentially, they provide analytic capabilities but without the storage or result delivery features needed for a complete BAD platform.

\noindent{\bf Data-centric approaches}
The most closely related work to BAD has been on supporting continuous queries via a \emph{data-centric} approach, i.e., finding ways to treat user queries as ``just data'' rather than as unique data flows. To this end, NiagaraCQ \cite{NiagaraCQ} performed a live analysis of standing queries to detect ``group signatures,'' which are groups of queries that perform a selection over the same attribute and that differ only by the constant of interest (e.g., age=19 vs. age=25). Given these group signatures, it created a dataset of the constants used and incrementally joined this dataset with incoming data to produce results for multiple users via a single data join. The growing field of Spatial Alarms \cite{bamba2008scalable,lee2016road,thakur2015planetsense} serves to issue alerts to users based on objects that meet spatial predicates. Spatial predicates are directly stored as objects (data) in an R-Tree, and incoming updates are then checked against all of the standing queries by simply performing a spatial join with this R-Tree.
 
The technical approach taken by NiagaraCQ and Spatial Alarms of treating continuous queries as data is one of the main inspirations for our own subscription scaling work. Both systems had limitations that we seek to relax in our work. NiagaraCQ was designed to operate using a very limited query language designed for XML data. Spatial Alarms focused on one special use case (where the queries are locations) rather than on the problem as a whole. We build on these ideas for the more general world of Big Data, e.g., with horizontally partitioned data and a more fully expressive query language.

\section{Passive BDMS} \label{asterix}
Our aim here is to start with a passive BDMS, in this case AsterixDB, and then show how to transform it into a BAD system. We start with an introduction to some of the advantages that we inherit from AsterixDB as well as the limitations of a passive BDMS. This will serve as a preamble to creating our Active Toolkit (Section \ref{active-asterix}).

\begin{figure}[!ht]
  \centering
  \includegraphics[width=0.45\textwidth]{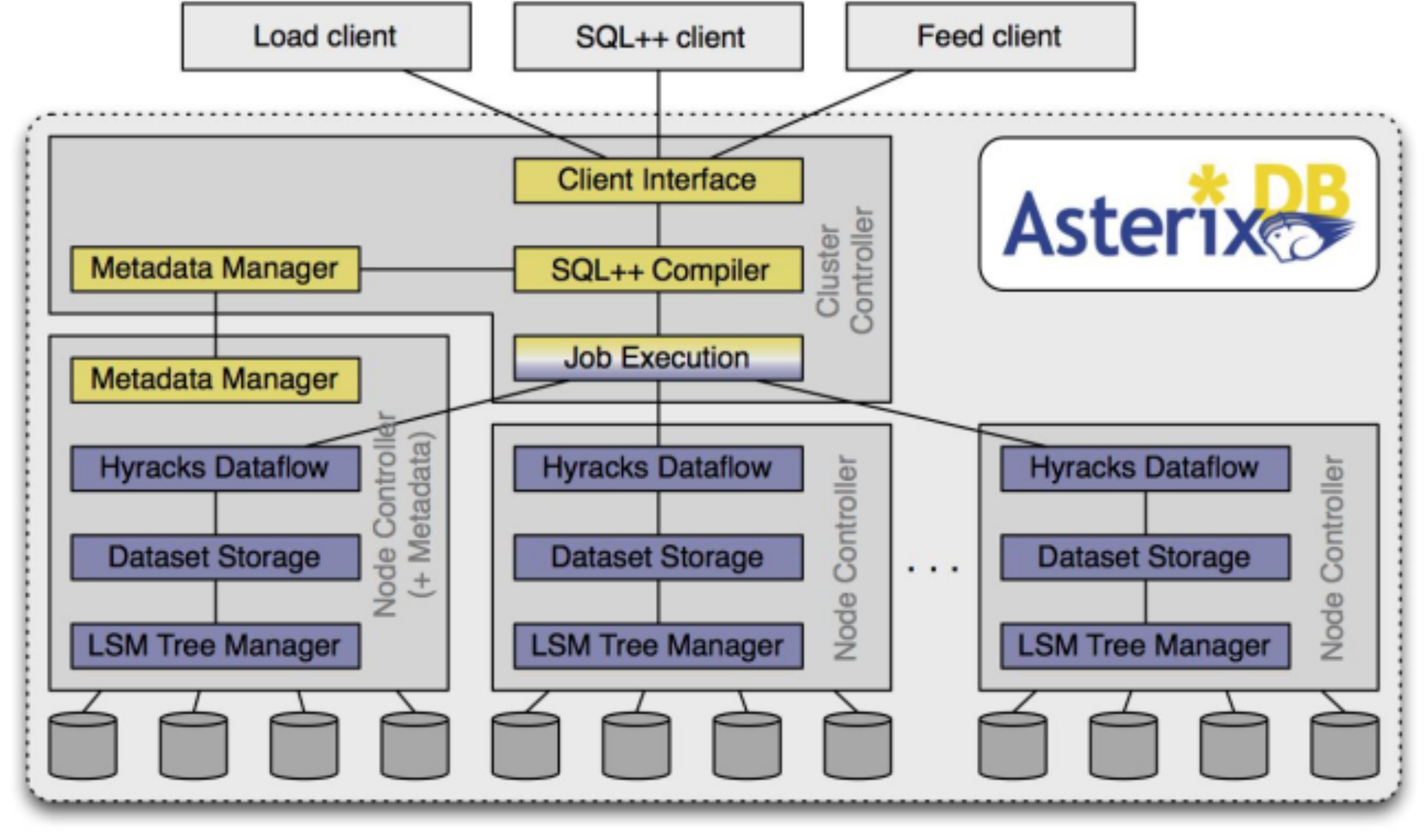}
    \caption{The architecture of passive AsterixDB}
  \label{asterix-architecture}
\end{figure}

Apache AsterixDB (see Figure~\ref{asterix-architecture}) is a full-featured BDMS that supports all of the prerequisites listed in Section \ref{requirements}. The underlying runtime engine for executing its queries, written in SQL++, is the Hyracks data-parallel platform~\cite{Hyracks-ICDE}. Queries are compiled and optimized via the Algebricks extensible algebraic parallel query planning and optimization framework \cite{AlgebricksSoCC}. AsterixDB has fully developed support for rich Big Data types, including GeoJSON and other advanced spatial types, which fits well with the location-oriented applications we are considering~\cite{deem2018}. Another feature of AsterixDB that makes it particularly suitable for becoming active is the provision of \emph{data feeds} built on top of LSM (Log-Structured Merge) tree storage technology, allowing for fast data ingestion~\cite{alsubaiee2014storage,grover2015data,chenlsm19,DBLP:journals/pvldb/WangC19}.

\begin{figure}
\begin{minipage}[t]{0.5\textwidth}
{\small
\begin{sql}
CREATE DATAVERSE emergencyNotifications;
USE emergencyNotifications;

CREATE TYPE UserLocation AS {
	location: circle,
	userName: string,
	timestamp: datetime
};
CREATE TYPE EmergencyReport AS {
	reportId: uuid,
	Etype: string,
	location: circle,
	timestamp: datetime
};
CREATE TYPE Contact AS {
    contactName: string,
    phone: int64,
    address: string?
};
CREATE TYPE EmergencyShelter AS {
	shelterName: string,
	location: point,
	contacts: {{ Contact }}?
};

CREATE DATASET UserLocations(UserLocation)
    PRIMARY KEY userName;
CREATE DATASET Shelters(EmergencyShelter)
    PRIMARY KEY shelterName;
CREATE DATASET Reports(EmergencyReport)
    PRIMARY KEY reportId autogenerated;

CREATE INDEX location_time ON UserLocations(timestamp)
    TYPE BTREE;
CREATE INDEX u_location ON UserLocations(location) 
    TYPE RTREE;
CREATE INDEX s_location ON Shelters(location) 
    TYPE RTREE;
CREATE INDEX report_time ON Reports(timestamp)
    TYPE BTREE;
\end{sql}
\begin{center}
{\small \textbf{(a)}}
\end{center}
\begin{sql}
SELECT report, u.userName FROM
 (SELECT VALUE r FROM Reports r 
    WHERE r.timestamp > 
    current_datetime() - day_time_duration(``PT10S'')
  ) report,
UserLocations u
WHERE spatial_intersect(report.location,u.location);
\end{sql}
\begin{center}
  \vspace{-0.3cm}
{\small \textbf{(b)}}
\end{center}
\begin{sql}
INSERT INTO Shelters (
{``shelterName'' : ``swan'' , 
 ``location'' : point(``2437.34,1330.59'') ,
 ``contacts'' : {{
	{ ``contactName'' : ``Jack Shepherd'',
		``phone'' : 4815162342 },
	{ ``contactName'' : ``John Locke'',
		``phone'' : 1234567890 }
	}}}
);
\end{sql}
\begin{center}
  \vspace{-0.4cm}
{\small \textbf{(c)}}
\end{center}
  \vspace{-0.4cm}
\caption{Examples of (a) ADM data types, datasets, and indexes, (b) a SQL++ query, and (c) a SQL++ INSERT statement}
\label{aql-example}
}
\end{minipage}
\end{figure}

Figure \ref{aql-example} illustrates by example the AsterixDB data model (ADM) and language; part (a) shows the data type, dataset, and index definitions that could be used for our example application, as well as a SQL++ {\tt SELECT} query and a SQL++ {\tt INSERT} statement.
The query in part (b) finds the emergencies that have been reported in the last ten seconds and joins them spatially with the locations of users, and part (c) shows how a new shelter could be added.

When a request (e.g. the {\tt SELECT} query in Figure \ref{aql-example}b) is sent to AsterixDB, it is first parsed and optimized into an algebraic parallel query plan. This plan is then physically compiled into a Hyracks \emph{job}, a directed acyclic operator/connector graph (DAG), that is distributed to the cluster to execute. A high-level DAG will be seen in Figure \ref{channeldag}.

\subsection{Limitations of Passive BDMS} \label{limitations}
AsterixDB was architected with Big Data capabilities in mind; however, it has some limitations from the perspective of the needs of an active framework. With the exception of \emph{data feeds}, every job performed is tied to an explicit user interaction, from start to finish. Compounding this problem is the fact that jobs in AsterixDB are treated in isolation. Consider our use case where users want to know about emergencies near them as they occur. In passive AsterixDB, information is only gained by directly requesting it (e.g., running a query to check recent emergencies near the user's current location). If a user wanted to continuously check for new information, the user would need to continuously request it (e.g., by sending a new request every 10 seconds to check for new emergencies from the last 10 seconds). This could be done in the following way:

\begin{enumerate}
    \item The user sets up a cron job that runs every 10 seconds and calls the AsterixDB REST API.
    \item At each execution, the script sends a query to AsterixDB.
    \item AsterixDB treats this request as a new (never-before-seen) job, which must be parsed, compiled, and optimized. Then it is distributed to the nodes of the cluster.
    \item The job for the query is finally executed.
    \item AsterixDB performs job cleanup, including the removal from all nodes of the information for the job.
    \item Steps (2-5) are repeated ad infinitum. 
\end{enumerate}

This query model works well for a query that is run once, but clearly becomes wasteful when a job is repeated, resulting in significant shortcomings: 

\begin{enumerate}
\item  The work for steps (3) and (5) is repeated every ten seconds, even though it is exactly the same every time.
\item  Every execution of the job requires explicit triggering by an outside source (the cron job in step (2)).
\item Both (1) and (2) are multiplied by the number of users who are performing the same task in parallel.
\end{enumerate}

\section{The Active Toolkit} \label{active-asterix}
To overcome the above limitations we created an \emph{Active Toolkit} for AsterixDB. It contains
four tools needed to build a BAD framework, namely:

\begin{enumerate}
\item  \emph{Data feeds} to rapidly ingest application data. A data feed represents the flow of scalable rapid data into one or more datasets.
\item  \emph{Deployed jobs} that can perform arbitrary SQL++ tasks. They get compiled and distributed once and used multiple times.
\item  \emph{Data channels} to actively process data with respect to subscriber interests. A single channel is compiled once and shared by a scalable number of users yet produces individualized staged results.
\item  \emph{Procedures} to use deployed jobs to perform other active management processes regularly and efficiently.
\end{enumerate}

Returning to Figure \ref{sys}, these four tools realize the vision of a BAD Platform. The Data Feeds enable large numbers of Data Publishers to provide rapid data to the BAD Platform. Deployed Jobs and Procedures enable Data Managers to maintain and monitor the BAD Platform. Most importantly, Data Channels provide the mechanism for taking the data as a whole and producing enriched individualized results for Data Subscribers, which the Broker Network can then deliver.

\subsection{Data Feeds} \label{feeds}

Since data in an active environment is being generated rapidly, it would not be efficient to insert records one by one through a typical DDL statement. Alternatively, bulk data loading is useful when there is a large collection of new data sitting on the disk waiting to be imported, but for data incoming as a continuous stream, we need a different mechanism. Data feeds \cite{grover2015data} were initially introduced as a new feature in AsterixDB to persist continuous data streams into AsterixDB datasets. Starting from there, we have made a series of updates to data feeds in order to make them even more effective for BAD scenarios.

In \cite{grover2015data}, we introduced the notions of a ``primary feed'', which gets data from an external data source, and ``secondary feeds'' that can be derived from another feed. In addition, either/both could have an associated user-defined function (UDF). 
Data feeds enable users to attach UDFs onto the feed pipeline so that the incoming data can be annotated (if needed) before being stored.
A user could use that architecture to build a rich ``cascade network'' that routes data to different destinations for particular purposes. 

\begin{figure}[!ht]
  \centering
  \includegraphics[width=0.45\textwidth]{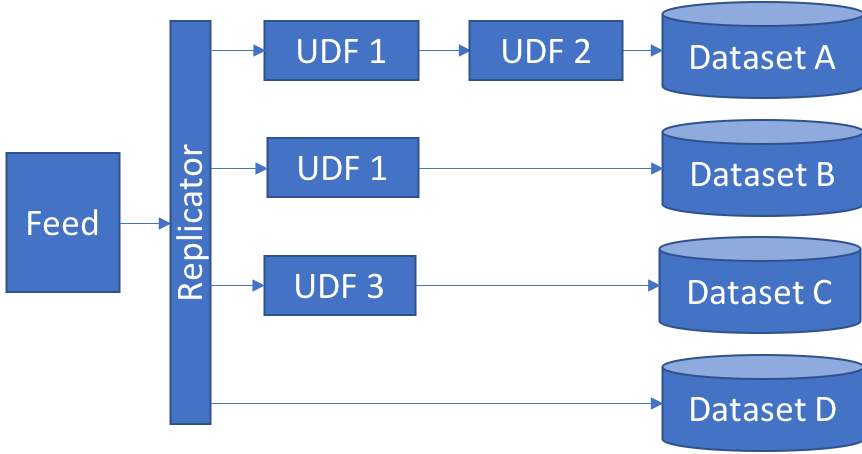}
    \caption{The updated feed dataflow}
  \label{new_feed}
\end{figure}

While that initial architecture introduced a lot of flexibility for building feed dataflow networks, it also brought extra overheads related to persisting the data and additional complexities in maintaining the dataflow network. In a BAD application, the timeliness of data and the robustness of the network outweigh the user-level flexibility of defining a complex feed network. To meet the BAD requirements, we have \textit{redesigned} the feed dataflow in a more succinct yet equally powerful architecture (in terms of the set of addressable use cases), as depicted in Figure \ref{new_feed}.

In the updated architecture, we have removed the previous cascade network and instead branch out sub-dataflows earlier with a ``replicator''. The sub-dataflows are isolated from one other so that the data movement in each sub-dataflow can proceed without interfering with the others. The UDFs attached to each path are evaluated separately as well.

A feed (on the far left in Figure \ref{new_feed}) internally consists of an adapter and a parser. The adapter gets data from an external data source, and the parser translates the incoming data into ADM objects. Feeds were introduced with Socket, File, RSS, and Twitter adapters as well as JSON, ADM, Delimited Data, and Tweet parsers to handle common use cases. In building the BAD platform, we soon realized the need for adding many more adapters and parsers in order to handle a larger selection of data sources and formats. To address this issue we created a pluggable adapter and parser framework so that users can add their own parsers and adapters and use them 
just like the native ones.

In~\cite{grover2015data}, data feeds provided {\tt INSERT} semantics since they were initially intended to be continuously-running sources of new data. Later experiences, including using feeds for ``BAD'' applications, led us to
add {\tt UPSERT} (i.e., insert if new, else replace) semantics as an option as well. Incoming data may in some cases contain duplicates (e.g., the same Tweet arriving via an ``at least once'' connection, or the same emergency report from several agencies).  In other  cases, users may explictly intend for the stream to contain updates, and they may only want to keep the latest information (e.g., users' current locations).  {\tt UPSERT} semantics are in fact the new default for feeds.

\begin{figure}
\begin{minipage}[t]{0.5\textwidth}
{\small
\begin{sql}
USE emergencyNotifications;

CREATE TYPE UserLocationFeedType AS {
	location: circle,
	userName: string
};
CREATE TYPE EmergencyReportFeedType AS {
	Etype: string,
	location: circle
};

CREATE FEED LocationFeed WITH
{
    ``adapter-name'' : ``socket_adapter'',
    ``sockets'' : ``bad_cluster.edu:10009'',
    ``address-type'' : ``IP'',
    ``type-name'' : ``UserLocationFeedType'',
    ``format'' : ``adm''
};
CREATE FEED ReportFeed WITH
{
    ``adapter-name'' : ``socket_adapter'',
    ``sockets'' : ``bad_cluster.edu:10008'',
    ``address-type'' : ``IP'',
    ``type-name'' : ``EmergencyReportFeedType'',
    ``format'' : ``adm''
};
\end{sql}
\caption{Create data feeds for the Reports and UserLocations}
\label{feeds-ddl}
}
\end{minipage}
\end{figure}

In our example application, we assume that the data being ingested into UserLocations and Reports are highly dynamic, as the user locations are being updated and reports are being generated frequently. 
Figure \ref{feeds-ddl} depicts feeds being created for both datasets.
Both feeds in this example expect data in ADM format. The default create-feed statement creates a feed with {\tt UPSERT} semantics.
The DDL demonstrates a socket adapter on a designated host (e.g., ``bad\_cluster.edu: 10008''). When clients come, they can connect to these endpoints and send their data directly. 

Note that Figure \ref{feeds-ddl} defines two additional datatypes, ``UserLocationFeedType'' and ``EmergencyReportFeedType'', for our feeds. Incoming data from publishers is not required to have a timestamp, thus the datatype for the incoming data does not have a timestamp. For a BAD application, however, the timestamp is an important field as it will be used later for generating the emergency notifications. 
To annotate the incoming data with proper timestamps, we create a UDF and attach it to a feed so that the incoming data is timestamped before it reaches the dataset (BAD nodes should be synchronized using NTP). We first create a function to add insert time, as shown in Figure \ref{add_insert_time}.  This function utilizes the built-in SQL++ functions ``current\_datetime()'' and ``object\_merge()'' to add a new field with the current timestamp to an incoming record, thus converting a record of the ``EmergencyReportFeedType'' into a record of the actual datatype, ``EmergencyReport.''

\begin{figure}
\begin{minipage}[t]{0.5\textwidth}
{\small
\begin{sql}
USE emergencyNotifications;
CREATE FUNCTION add_insert_time(record) {
    object_merge({``timestamp'': current_datetime()}
    , record)
};

/*
Sample Incoming Record:
{``Etype'' : ``storm'', 
``location'' : circle(``846.5, 2589.4, 100.0'')}
Sample Output Record:
{``Etype'' : ``storm'', 
``location'' : circle(``846.5, 2589.4, 100.0''), 
``timestamp'' : datetime(``2018-08-27T10:10:05'')}
*/
\end{sql}
\caption{Create the ``add\_insert\_time'' function}
\label{add_insert_time}
}
\end{minipage}
\end{figure}

As the final step in setting up a data feed, we attach the UDF to the feed pipeline, connect the feed to the dataset, and start the feed. The DDL statements to accomplish this are shown in Figure \ref{connfeeds-ddl}. All incoming records for the UserLocations and Reports datasets will now be annotated with an arrival timestamp that will be used shortly in their associated \emph{data channels}.

\begin{figure}
\begin{minipage}[t]{0.5\textwidth}
{\small
\begin{sql}
USE emergencyNotifications;

CONNECT FEED LocationFeed TO DATASET UserLocations 
    APPLY FUNCTION add_insert_time;

CONNECT FEED ReportFeed TO DATASET Reports 
    APPLY FUNCTION add_insert_time;

START FEED LocationFeed;
START FEED ReportFeed;
\end{sql}
\caption{Connect the data feeds to both datasets with function}
\label{connfeeds-ddl}
}
\end{minipage}
\end{figure}

\subsection{Deployed Jobs} \label{deployedjobs}
The overhead of parsing, compiling, optimizing, and distributing a job (e.g., an ``INSERT'' or ``QUERY'' execution pipeline) can be especially time-consuming for small jobs. For example, fetching a single record by primary key in AsterixDB currently takes around 20 milliseconds regardless of the size of the data cluster.
This is because the process of parsing, compiling, optimizing, and distributing a job incurs a penalty of 10-20 milliseconds before the job even starts executing. This is ``noise'' for longer-running Big Data analytics queries, but for small jobs this process can become dominant. 

We created an extendable class of jobs called \textit{deployed} jobs to address this overhead. A deployed job is a new first-class citizen in the runtime of AsterixDB that is created once but can be run many times. Deployed jobs are roughly similar in function to the prepared query facilities found in many conventional relational database systems. Deployed jobs can be created for the following types of SQL++ tasks: {\tt DELETE}, {\tt INSERT}, and {\tt SELECT} (query). 

\begin{figure}[!ht]
  \centering
  \includegraphics[width=0.45\textwidth]{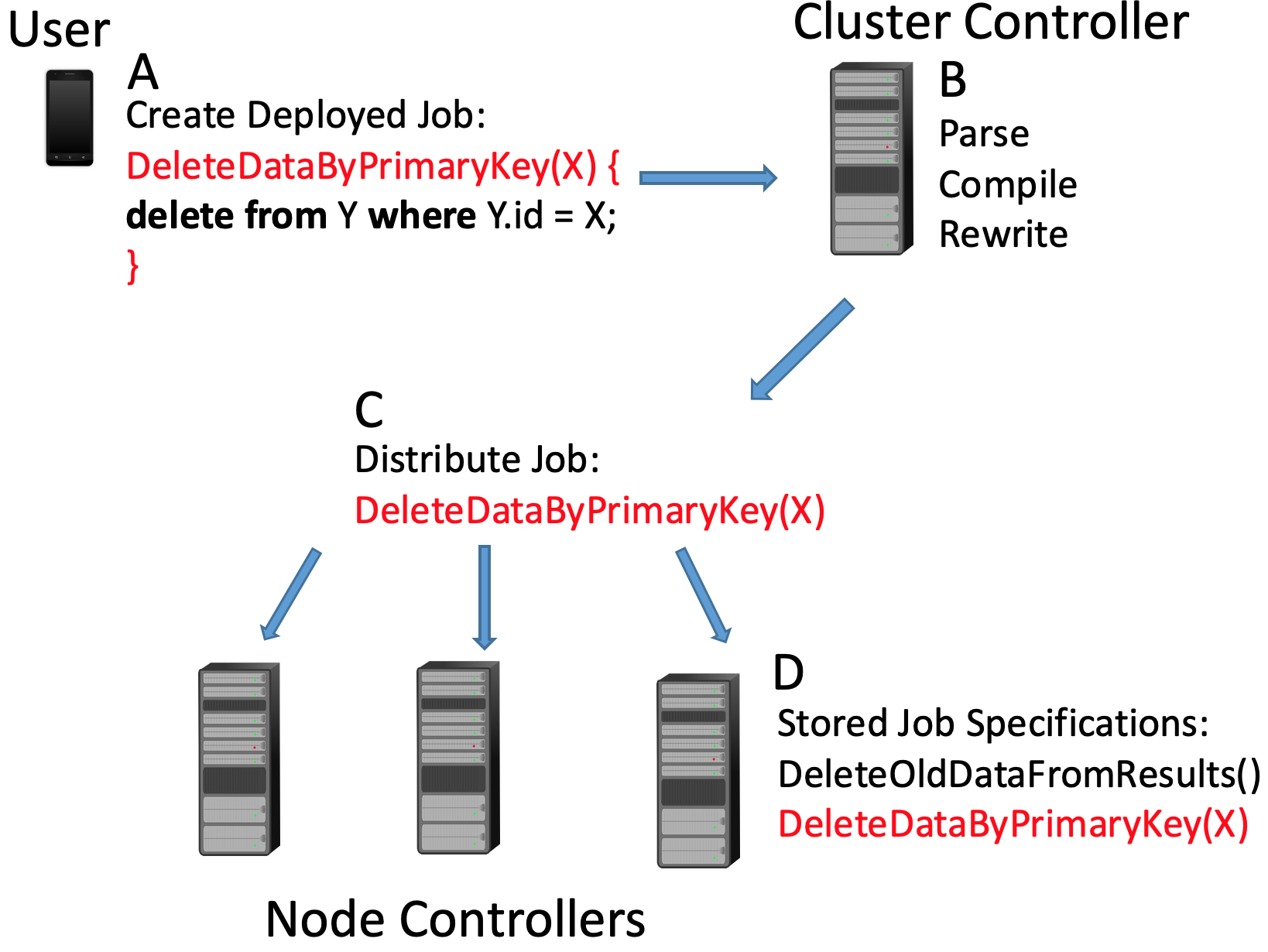}
    \caption{Deploying a Job}
  \label{jobdeployment}
\end{figure}

When a deployed job is created (see Figure \ref{jobdeployment}), the SQL++ syntax for the job is provided (see step A). This is then parsed, compiled, and optimized once to produce a job specification (B). The resulting job spec is then distributed (C) and cached at each node (D).

\begin{figure}[!ht]
  \centering
  \includegraphics[width=0.45\textwidth]{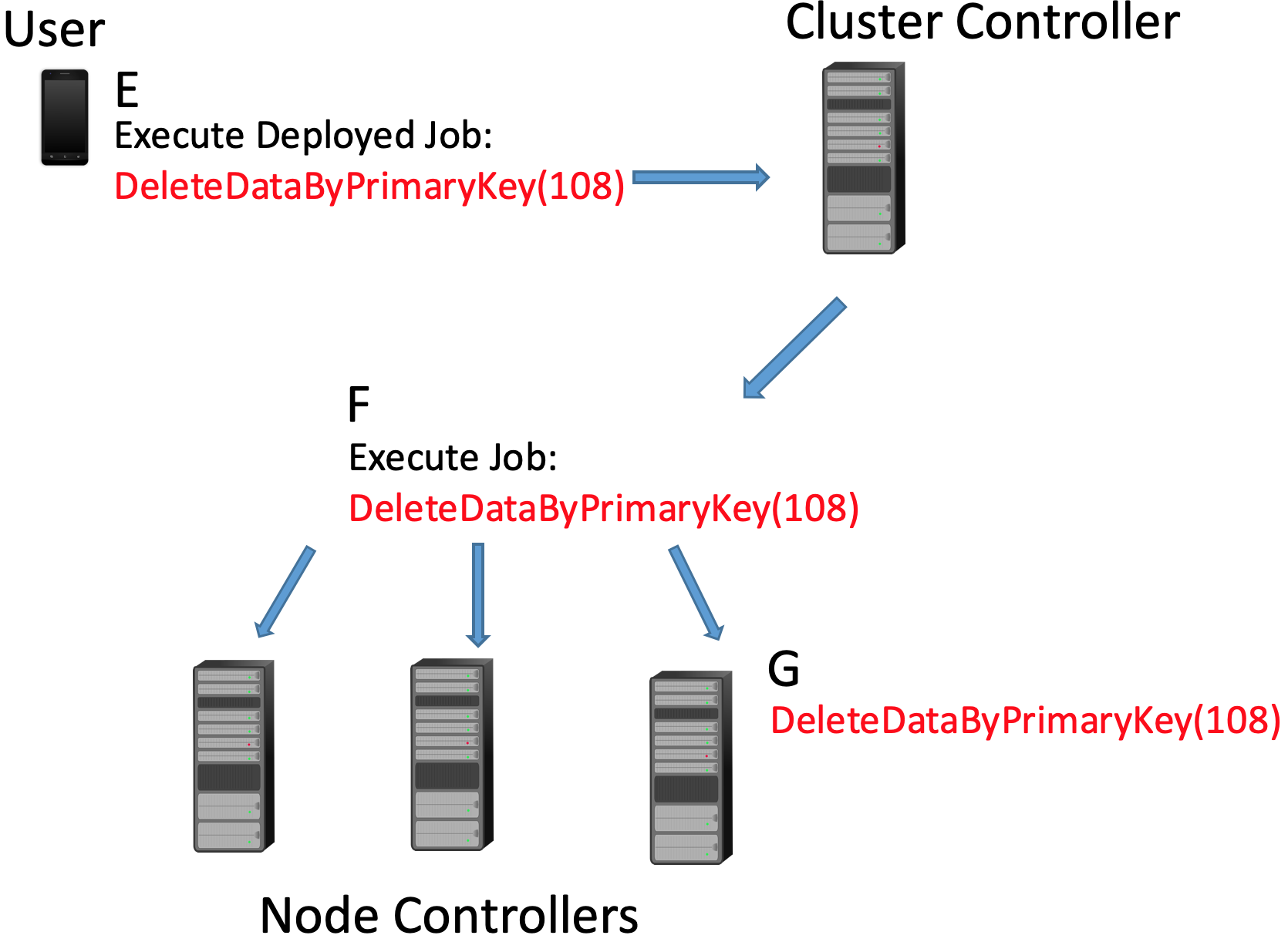}
    \caption{Executing a Deployed Job}
  \label{jobexecution}
\end{figure}

When executing a deployed job (Figure \ref{jobexecution}), it is simply referenced by name (E). The cluster controller sends an ``execute'' command (F) to the nodes, which then execute the job (G) using the stored job spec.

Note that the deployed job in Figures \ref{jobdeployment} and \ref{jobexecution} has a parameter representing the primary key of the record to be deleted. This parameterization is another enhancement that makes deployed jobs more robust. There are cases where different users may want to run a similar job that differs by only some set of parameters (in this case the id of the record). To support this, we implemented \textit{parameterized} deployed jobs. The parameter values are passed to the node controller for the given job execution by the cluster controller. At runtime, an added operator in the job's ongoing plan fetches the value for a given parameter. At job cleanup, the stored parameter values are removed for the job. Allowing users to share parameterized jobs will be examined further in the channels and the procedures.

A deployed job is one limited special case of an active job. More generally, a BAD platform should be able to support a scalable number of users (through a simple interface) who subscribe to data of interest to them. This implies actively processing data as it changes, storing new results as they are found, and delivering them to the data subscribers. We achieve this through the next feature of the Active Toolkit: \emph{repetitive data channels}.

\subsection{Channels} \label{channels}

\subsubsection{Channels for users}

We introduce the notion of a channel as a new, scalable mechanism that allows users to subscribe to data.
A channel is a shared deployed active job that produces individualized data for subscribing users. In order to scale, a channel is implemented as a parameterized query with each user specifying their individual parameter values of interest.

Consider our example application, where users want to be notified when emergencies occur that intersect with their current locations. A natural implementation using passive AsterixDB would be through \textit{polling}. Every user would explicitly poll the data cluster, at some interval, to see whether something new has occurred since the last poll. This would incur a steep penalty since every instance of every poll would be seen and compiled as a brand new query. We examine the performance of such a passive approach in Section \ref{experiments}.

AsterixDB already provides an interface (\emph{functions}) for defining a passive parameterized query that polling users could utilize. The move from passive to active for users can be colloquialized as follows: ``Rather than calling this function myself to check for data, I would like the function to call me when there is data of interest to me.'' Or, more succinctly, ``You've got data!''

A \emph{repetitive data channel} can be thought of as an active, shared version of a function (in fact the channel DDL makes use of the existing SQL++ function DDL) that utilizes an optimized deployed job to leverage shared interests but that produces individualized results for many users at once based on their individual parameters and sends notifications when new data is produced.

\begin{figure}
\begin{minipage}[t]{0.5\textwidth}
{\small
\begin{sql}
USE emergencyNotifications;

CREATE FUNCTION RecentEmergenciesNearUser(userName) {  
  SELECT report, shelters
  FROM 
    (SELECT VALUE r FROM Reports r 
      WHERE r.timestamp > current_datetime() - 
        day_time_duration(``PT10S")) report,
    UserLocations u, 
    (SELECT s.location FROM Shelters s WHERE
      spatial_intersect(s.location,u.location)) shelters
  WHERE u.userName = userName
    AND spatial_intersect(report.location,u.location)
};
\end{sql}
\begin{center}
  \vspace{-0.3cm}
{\small \textbf{(a)}}
\end{center}
\begin{sql}

RecentEmergenciesNearUser(``dharma1'');
\end{sql}
\begin{center}
  \vspace{-0.3cm}
{\small \textbf{(b)}}
\end{center}
\caption{DDL (a) for a function that finds recent emergencies near a given user, and an example invocation (b) of the function}
\label{function}
}
\end{minipage}
\end{figure}

We provide an SQL++ DDL extension for channels that leverages AsterixDB parameterized function definitions. As an example, recall the query in Figure \ref{aql-example} that joined recent emergency reports with the UserLocations dataset. Suppose that we want to create a function that will run a similar query on behalf of a single user. We can see such a function in Figure \ref{function}(a). When a user calls RecentEmergenciesNearUser(``dharma1'') in Figure \ref{function}(b), the variable ``userName'' will be replaced with ``dharma1'' in the query, and then the query will be treated normally. This provides a nice way to describe exactly the type of shared query that users of our example application would want to run. 
Note that the query in Figure \ref{function}(a) also enriches (personalizes) the user's results with nearby shelter information.

\begin{figure}
\begin{minipage}[t]{0.5\textwidth}
{\small
\begin{sql}
USE emergencyNotifications;

CREATE REPETITIVE CHANNEL EmergenciesNearMe USING 
RecentEmergenciesNearUser@1 PERIOD duration(``PT10S'');

CREATE BROKER BADBrokerOne AT ``BAD_broker_one.edu'';

SUBSCRIBE TO EmergenciesNearMe(``dharma1'') 
ON BADBrokerOne;
SUBSCRIBE TO EmergenciesNearMe(``johnLocke'') 
ON BADBrokerOne;

\end{sql}
\caption{DDL to create a channel using the function RecentEmergenciesNearUser@1, DDL for creating a broker, and DDL for creating a subscription to the channel}
\label{channel_ddl}
}
\end{minipage}
\end{figure}

Figure \ref{channel_ddl} shows how a channel can be created based on the function from Figure \ref{function}. Creating a repetitive channel requires two parts: a function for the channel to use and a time (repeat) period. Creating a channel will compile the query contained in the function into a single deployed job that then will be run repetitively based on the period provided (in this case every 10 seconds). Every time this deployed job is run it will produce a set of individualized results for all of the data subscribers.

It is worth noting that a trade-off of sharing a single channel execution is that users of the channel will also share the period of the channel (in this example 10 seconds). It may be the case that some users would desire the same sort of query behavior but with a different rate of analysis and delivery (e.g., ``send me the list of emergencies every hour''). Rather than making the performance match the user with the fastest demands, and thereby performing work more often than necessary for other users, multiple channels can be created with different periods (e.g., a 10-second channel and a 1-hour channel) to enable more capabilities for users. Of course this will also come with the cost of running multiple channels in parallel.

In addition to the channel, Figure \ref{channel_ddl} also shows how to create a \emph{broker} as a recognized subscription endpoint in BAD AsterixDB. In order to provide scalability, data subscribers connect to the cluster through BAD brokers, providing a one-to-many connection to BAD AsterixDB (brokers are discussed in more detail in Section \ref{threelayers}). When a data subscriber subsequently subscribes to a channel, the broker acting on behalf of the subscriber will provide: (i) the parameters relevant for that subscriber (in this case the id of the user), and (ii) the name of the broker making the request (in this case \emph{BADBrokerOne}). Once a subscription has been created, the subscriber will begin to receive results for emergencies near her changing location over time. 

\begin{figure}[!ht]
  \centering
  \includegraphics[width=0.45\textwidth]{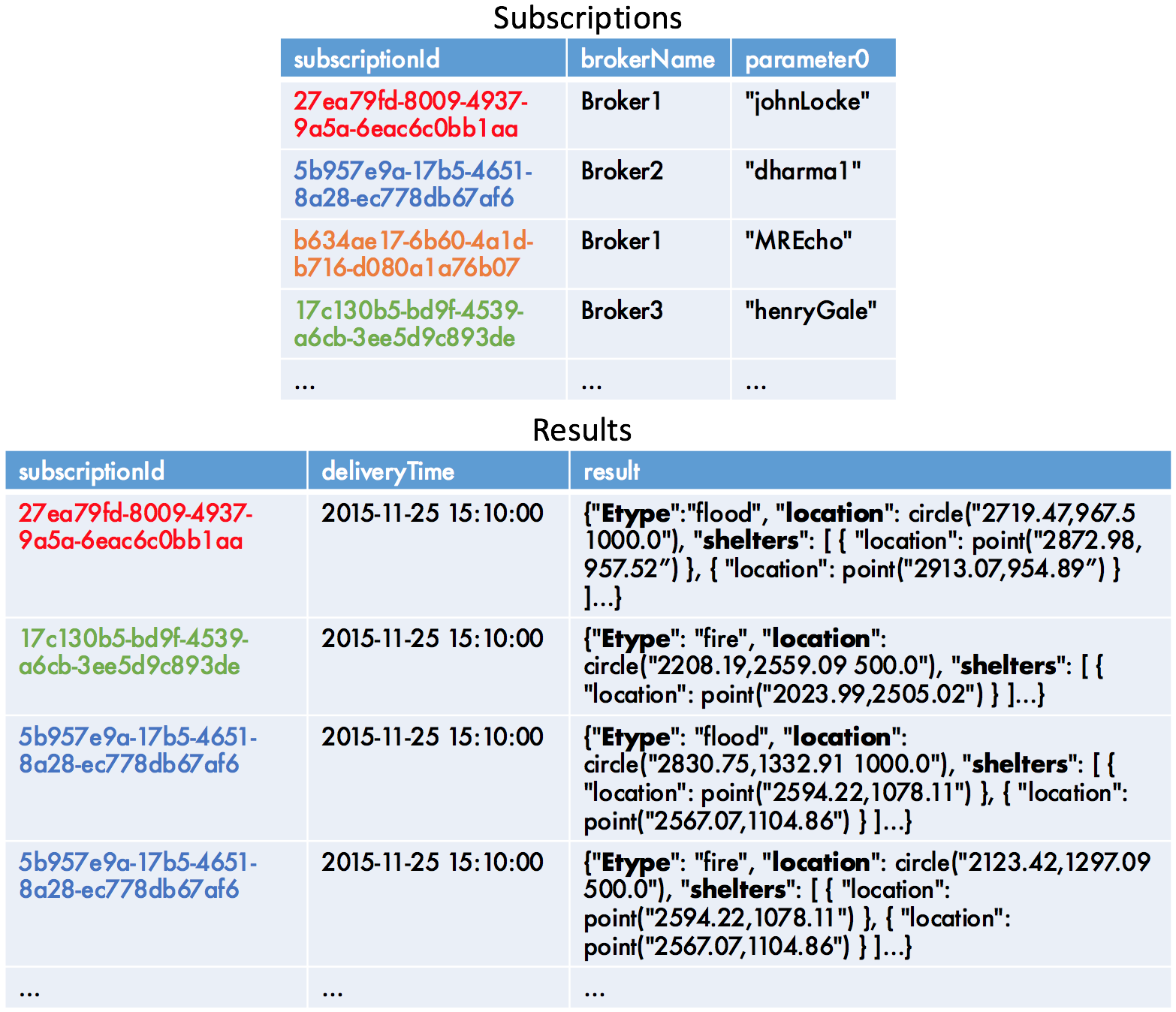}
    \caption{The subscription and results tables for the EmergenciesNearMe channel}
  \label{channeltables}
\end{figure}

\subsubsection{Channels under the hood}

We now discuss in detail how BAD AsterixDB creates and manages data channel work flows under the hood using an \emph{EmergenciesNearMe} channel as an example. When this channel is created, two new internal datasets will be created: \emph{EmergenciesNearMeSubscriptions} and \emph{EmergenciesNearMeResults}. The subscriptions dataset contains one record for each subscription that has been created. This includes three important pieces of information: (a) an automatically generated id for the subscription, (b) the name of the broker servicing the subscriber of the subscription, and (c) the channel parameter values for the subscription. The results dataset is where the result records for the channel will be stored (including their subscription ids). An example of these tables appears in Figure \ref{channeltables}.
Although these tables will start out empty, they are depicted with data to illustrate how their data could look over time. There are four subscriptions shown, along with results produced for some of these subscriptions.

\begin{figure}[!ht]
  \centering
  \includegraphics[width=0.45\textwidth]{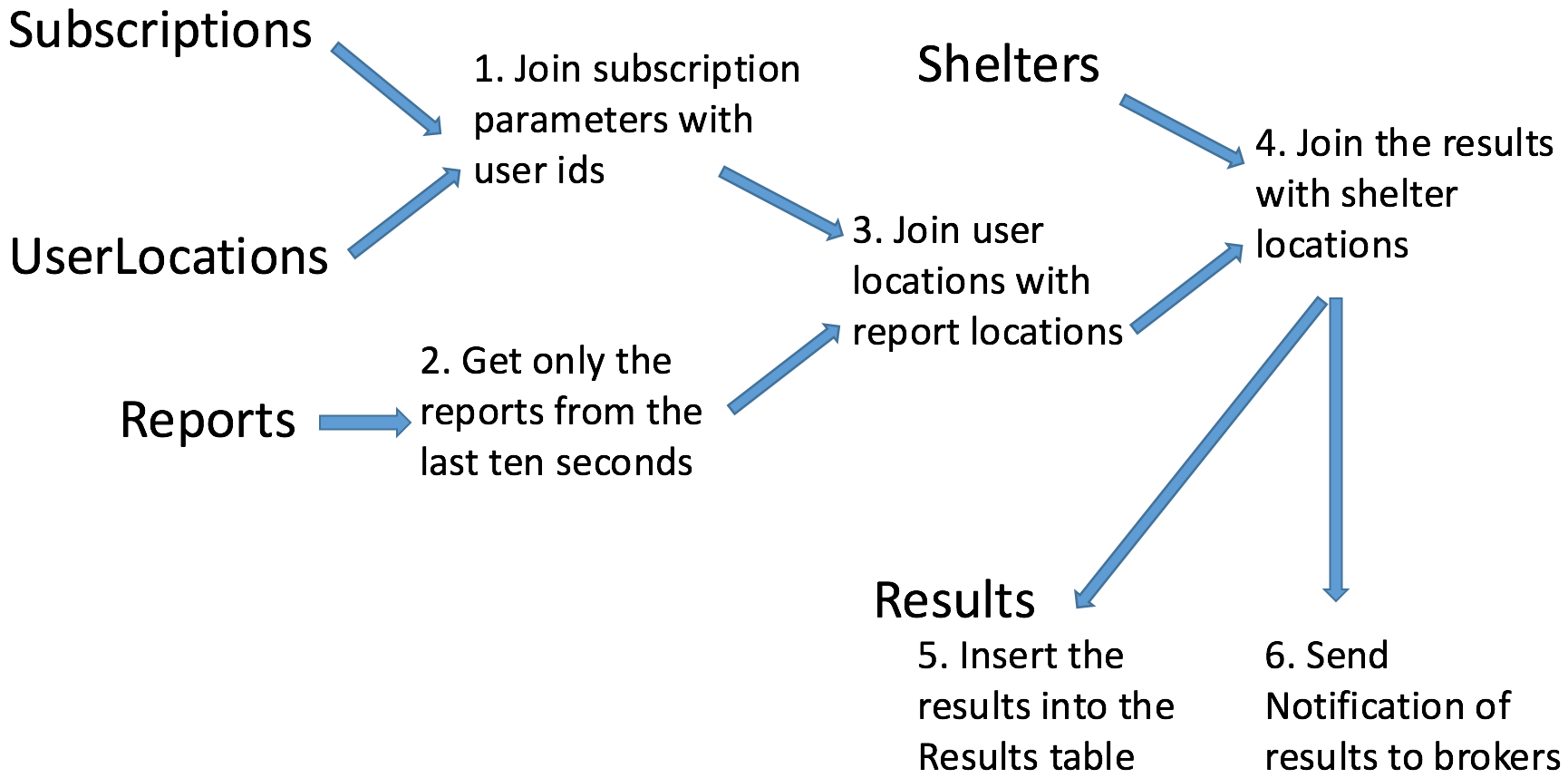}
    \caption{The deployed job for executing the channel EmergenciesNearMe}
  \label{channeldag}
\end{figure}

Once these two tables have been created, the channel will be compiled, optimized, and distributed to the node controllers. 
Rather than running the function separately for each subscription, a join is created between the function body and the \emph{EmergenciesNearMeSubscriptions} dataset on the parameter values. The results produced are inserted into the \emph{EmergenciesNearMeResults} dataset. This job is then optimized into a plan DAG by AsterixDB's rule-based job optimizer (see Figure \ref{channeldag}) that includes the following steps:
\begin{enumerate}
    \item Join the Subscriptions and UserLocations datasets to find the locations of the subscribers.
    \item Utilize the time index of Reports to fetch only the recent Reports (last ten seconds, from the function query).
    \item Perform a spatial join between the results of steps (1) and (2).
    \item Enhance the result with the nearby shelters (also spatially joined).
    \item Insert the results into the results table.
    \item Send the brokers notifications that new results have been created. 
\end{enumerate}

It is important to note the advantages of this approach over a passive polling method. If users were polling individually, each poll request would produce an individual job very similar to Figure \ref{channeldag}, but with the \emph{Subscriptions} table being replaced by a single input value (the id of the user making the request) and with the \emph{results} being delivered directly back to that user. In contrast, the deployed channel job executes once every ten seconds (the period of the channel) on behalf of all subscribers, potentially produces new results, and requires no intervention from the users. Section \ref{threelayers} discusses in more detail how the results (and the subscriptions) are communicated from the data cluster to each subscriber (and vise-versa).

\subsubsection{The Case for Continuous Channels} \label{continuous}
Repetitive channels have the limitation that they rely on some fixed time interval to execute (e.g., ten seconds). While this is fine if data is produced and desired at a specific rate, it cannot handle two extreme but common cases: users wanting data at the moment of its creation (not waiting until the next channel execution), and events of interest occurring infrequently (not producing results for several executions, and thereby potentially wasting interim query processing resources). In this direction we are currently researching the next generation of channels, namely \emph{continuous channels}, inspired by \cite{goldberg1992using}, which will execute on data changes rather than relying on fixed execution periods.

It is interesting to note that when events are time-driven and the channel functions are time-qualified (as in our sample application), repetitive channels can provide a batch-y approximation to continuous channels, as they only execute on and produce a small set of data if the repeat interval is not too large.

\subsection{Procedures} \label{procedures}

Procedures are another entity built on top of deployed jobs to help maintain and provide tools for a BAD application. For example, brokers might want to retrieve lists of their current subscriptions to a given channel. Rather than having each broker send such a request as a new job each time, an application administrator could create the first example procedure (\emph{CountBrokerSubscriptions}) in Figure \ref{procedure_ddl} that can then be used multiple times by multiple brokers. This also shows how one can provide a parameter to a procedure (in this case the name of the broker of interest). The value of the parameter is then passed when ``execute'' is called. Recall the execution pipeline in Figure \ref{jobexecution}. Roughly speaking, procedures are like a time-based version of the stored procedures found in the relational world.

In order to accomplish active objectives using procedures (addressing the limitations discussed in Section \ref{limitations}), we have augmented the deployed job capabilities by allowing users to specify an execution frequency when running a deployed job (e.g., 24 hours), thus allowing the creation of \emph{repetitive procedures}. Here the user will only make one explicit call. Subsequent executions will then happen actively, with no user interaction, every 24 hours.
Conceptually this can be seen as the simplest possible version of an active job. It can be noted that repetitive procedures can perform at scale many of the use-cases that triggers \cite{dayal1988hipac} were used for in traditional database systems, including inserting corollary information for newly inserted data and enforcing integrity constraints (albeit with a latency).

Managing a channel results dataset provides a use-case for such a repetitive procedure. The dataset can be thought of as a log of results being continually appended. This data might (depending on the type of application) be considered to be stale after some time threshold. In our example application, where users are notified of emergencies on an ongoing basis, we might only want to keep the old results in a broker-retrievable form for one day.

\begin{figure}
\begin{minipage}[t]{0.5\textwidth}
{\small
\begin{sql}
USE emergencyNotifications;

CREATE PROCEDURE CountBrokerSubscriptions(brokerName) {
    SELECT array_count(
        (SELECT sub 
        FROM EmergenciesNearMeSubscriptions sub 
        WHERE sub.BrokerName = brokerName))
};

EXECUTE CountBrokerSubscriptions(``BADBrokerOne'');

CREATE PROCEDURE deleteStaleResults() {
    DELETE result FROM EmergenciesNearMeResults
    WHERE result.channelExecutionTime < 
        current_datetime() - day_time_duration(``PT24H'')
} PERIOD duration(``PT24H'');

EXECUTE deleteStaleResults();

CREATE PROCEDURE SubCountsForEmergenciesNearMe(){
INSERT INTO SubscriptionStatistics (
    SELECT current_datetime() AS timestamp, b.BrokerName, 
        (SELECT VALUE array_count(
            (SELECT sub 
            FROM EmergenciesNearMeSubscriptions sub 
            WHERE sub.BrokerName = b. BrokerName)))
        AS subscriptions
    FROM Metadata.`Broker' b)
} PERIOD duration(``PT1H'');

EXECUTE SubCountsForEmergenciesNearMe();

\end{sql}
\caption{DDL for creating and executing three procedures (with the latter two being repetitive)}
\label{procedure_ddl}
}
\end{minipage}
\end{figure}

An application administrator can easily set up a procedure for cleaning up the results dataset using the DDL and DML for the second procedure (\emph{deleteStaleResults}) in Figure \ref{procedure_ddl}. The body of this procedure deletes channel results that are more than 24 hours old. Note that this procedure can be given an execution interval (24 hours). The ``execute'' call to initiate the active procedure will only need to be called once. The procedure will then continue to repeat every 24 hours. There are many other needs that procedures would be useful for in our application as well. For example, procedures could also be used to help evaluate broker utilization. The third procedure in Figure \ref{procedure_ddl} (\emph{SubCountsForEmergenciesNearMe}) will query the subscription counts for the \emph{EmergencyChannel} for every broker, on an hourly basis, and insert the results into a \emph{SubscriptionStatistics} dataset. Retrospective analytics can then be used on this dataset to tune the broker network itself.

There are many ways that procedures could enhance a BAD Platform, including: gathering statistics on the types of emergencies that are frequently producing results, finding the average number of results produced per execution, etc.

\begin{figure}[!ht]
  \centering
  \includegraphics[width=0.45\textwidth]{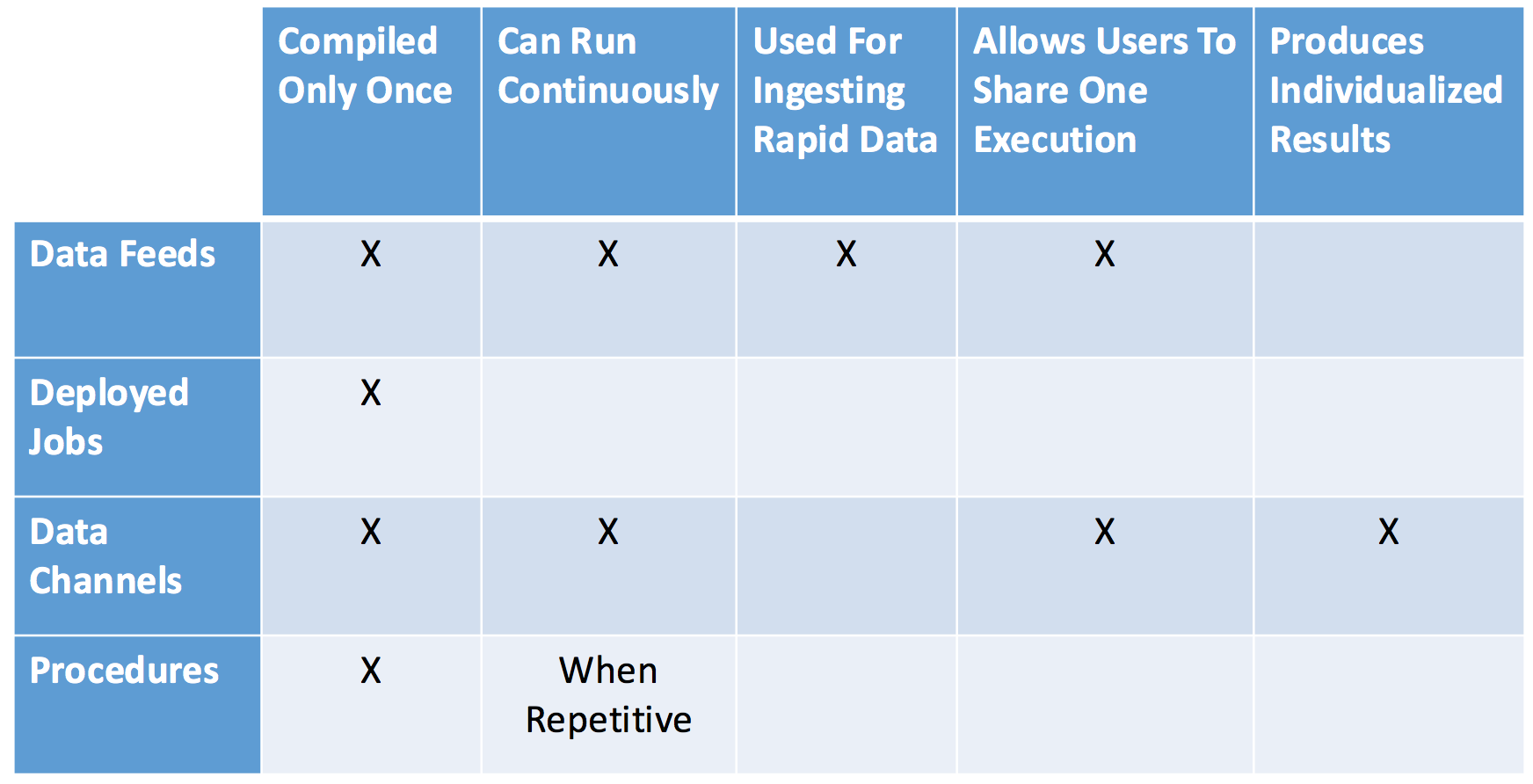}
    \caption{Comparison of Active Toolkit tools}
  \label{ActiveToolkit}
\end{figure}

The table in Figure \ref{ActiveToolkit} summarizes the abilities and differences of the tools in the Active Toolkit. Data Channels and Procedures are both extended implementations of deployed jobs.

\subsection{Users of the Active Toolkit} \label{usersRedux}

The term ``user'' could loosely apply to three different types of users in our example application, namely, Application Administrators, Data Publishers and Data Subscribers.

\subsubsection{Application Administrators}

An \textbf{Application Administrator} builds applications using the BAD framework. They have direct access to the data cluster for hosting datasets and {\em data feeds}. They have knowledge of: (1) Database Administration for the data stored in the Data Cluster for their applications, and, (2) the common interests of their ``users'' (eventual Data Subscribers). Based on user interests, an Application Administrator will create and manage parameterized {\em channels} that can be subscribed to in the application.

In our example scenario, the Application Administrator is who will create the \emph{emergencyNotifications} dataverse. She will then create the three datasets: \emph{Reports} (the continuously ingested reports), \emph{UserLocations} (the current location of each ``user'' (subscriber) of the application), and \emph{Shelters} (the relatively static metadata for shelter information, initially bulk loaded with all known shelters by the administrator).

The Application Administrator will then proceed to make this a BAD application by creating the {\em data feeds} for both \emph{Reports} and \emph{UserLocations} (DDLs shown in Figure \ref{feeds-ddl}), and by creating the subscribe-able repetitive channels for the application (via the DDL shown in Figure \ref{channel_ddl}). Lastly, she can create the relevant Procedures to help with the active management of the application (shown in Figure \ref{procedure_ddl}).

\subsubsection{Data Publishers}
{\bf Data Publishers} provide data in the form of streams of incoming records. These streams enter the Data Cluster directly via {\em data feeds}. In a typical use case, the data publishers will be exteral services that are known/trusted by the Application Administrators (such as news sites, social media data streams, or government agencies) and the incoming data will be broadly relevant to a given BAD application (e.g., emergency reports or weather broadcasts).

In our example scenario, the Application Administrator will provide the Emergency Report publisher with the cluster endpoint for sending reports to the \emph{ReportFeed}, and the publisher will then send its reports to this endpoint (e.g., ``bad\_data\_cluster.edu:10008'').

\subsubsection{Data Subscribers}
The third category is {\bf Data Subscribers}. They connect to BAD applications and subscribe to one or more of their channels. They are never given direct access to the data cluster, but instead perform all of their tasks via a nearby BAD Broker. This separation of the subscribers from the cluster provides several advantages. Rather than dealing with result data requests per subscriber, the cluster instead receives aggregated requests from brokers on behalf of many subscribers at once, and in the same way the cluster sends aggregate notifications to the brokers, rather than communicating with every subscriber. This limits the per-user impact on the cluster, freeing more resources for BAD tasks. In addition, this layered approach separates concerns, allowing brokers to focus on problems of result caching and communication issues with the subscriber, while the cluster deals with the data creation itself.  A {\em subscription} can be created for a specific channel and indicates the specific parameters of individual interest to a subscriber. Each subscription will be registered in the BAD data cluster with a subscription id. After its creation, the broker for a subscription will begin to receive new results from the channel that the given subscriber has subscribed to.

There are cases where a data subscriber may also serve as a data publisher. As an example, an application such as ours may want to allow users to provide (publish) their locations to the application (e.g., to enable subscriptions involving those locations). For such cases, an API is provided for a data feed that passes data from the subscribers through the brokers to the data cluster.

In our example application, users will want to subscribe to the \emph{EmergenciesNearMe} channel, so they will allow the application to have access to their current locations. The application will send this data via the brokers to the \emph{LocationFeed}. Users can then subscribe to the channel (their associated brokers will do so as shown in Figure \ref{channel_ddl}), at which point they will start receiving relevant results.

\subsection{A BAD Recap}
In this section, we presented the implementation of an active toolkit for AsterixDB that meets the \textit{BAD desiderata} proposed in Section~\ref{introduction}. 

Our main contributions in building the active toolkit can be summarized as follows:
\begin{itemize}
    \item We redesigned AsterixDB's \textbf{data feeds}, leading to a more succinct feed architecture that  branches out sub-dataflows with a ``replicator''. This allows data feeds to feed multiple datasets concurrently while maintaining high throughput. In addition, we introduced {\tt UPSERT}  semantics for data feeds involving ``at least once'' data sources; these are now the default semantics for feeds. 
    \item Inspired by prepared queries, we added \textbf{deployed jobs} into AsterixDB as the building block for repetitive channels and procedures. In addition to pre-compiling jobs to avoid per-query parsing, optimization, and job generation costs, we go further, deployng and caching the resulting compiled jobs on each node and invoking them with parameters to further reduce the per-job initiation cost.
    \item We introduced and implemented \textbf{channels}, a new mechanism that allows data subscribers to specify their data interests using parameters without having to write independent queries. Channels, which are defined functionally by application developers and built on top of deployed jobs, actively deliver (``push'') data of interest to subscribers instead of having them poll for data (``pull'') as in a passive data management system.
    \item We also introduced active \textbf{procedures} that enable administrators to easily manage the data life-cycle(s) in BAD applications by specifying data maintenance jobs to be executed on a pre-specified periodic basis. This periodic activation differentiates BAD procedures from the stored procedures typically found in the relational world.
\end{itemize}

With the active toolkit, \textbf{application developers} and \textbf{application administrators} can easily create and manage BAD applications to provide customized data notification and delivery services for \textbf{data publishers} and \textbf{data subscribers} using DDL statements based on a rich declarative query language. As we will discuss in Section~\ref{glue}, such a declarative and systematic solution saves administrators from the additional effort of gluing multiple systems together and orchestrating them to provide equivalent BAD functionalities. Additionally, with the optimizations employed in the BAD framework (efficient data ingestion, deployed jobs, etc.), BAD applications can provide such customized notification services efficiently at scale.

\section{BAD Layers} \label{threelayers}

\begin{figure}[!ht]
  \centering
  \includegraphics[width=0.5\textwidth]{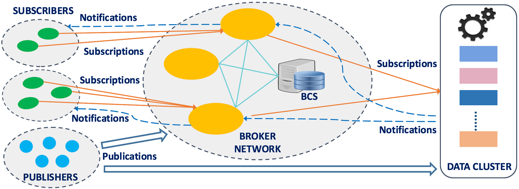}
    \caption{Communication in the BAD system}
  \label{layers}
\end{figure}

As shown in Figure \ref{layers}, there are three layers of communication for a BAD application. This provides a separation of concerns and allows each layer to perform tasks that are optimal for such a layer. Specifically, the Broker network is focused on efficiently handling both subscription communication and result delivery. Since the Broker network is structurally similar to a Pub/Sub Broker network, it can utilize state-of-the-art Pub/Sub scaling techniques and heuristics. On the other hand, the Data Cluster layer is built directly on a state-of-the-art Big Data Platform, and it can therefore capitalize on its capabilities (e.g., the distributed query engine). These details are discussed in further depth below. 

\subsection{Subscriber Network}
A user joining the application will first communicate with a Broker Coordination Service (BCS), which will assign a broker to communicate through. The BCS communicates with the brokers, and may change user assignments based on the loads of brokers and locations of users. If a broker fails, the user can also communicate with the BCS to receive a new broker assignment. The Application Administrator can decide to what level the users will be aware of the underlying channels and communications. For example, whether they are choosing from a list of channels to subscribe to or simply registering \emph{interests} using a higher-level interface. 

\begin{figure}[!ht]
  \centering
  \includegraphics[width=0.45\textwidth]{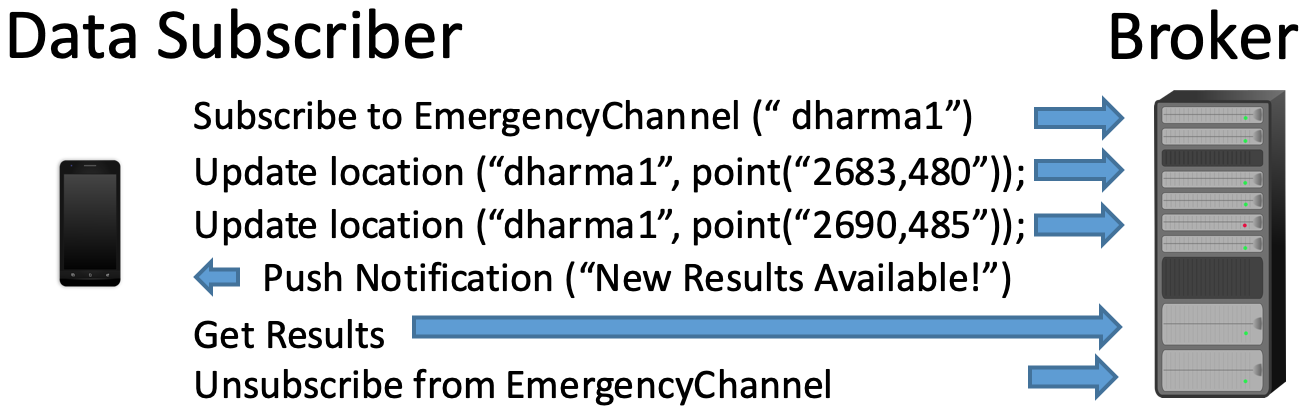}
    \caption{Communications between a subscriber's device and a broker}
  \label{subsriber-broker-API}
\end{figure}

The communications between a subscriber's device and a broker are as follows:

\begin{enumerate}
\item The user's application will send a \emph{subscribe} request to the broker (and will get back a subscription id that represents that subscription).
\item In the background, the application will continue to send the broker location updates.
\item When there are new results, the broker will send a \emph{push notification} to the application.
\item The application will send a \emph{get results} request to retrieve these results from the broker (including the subscription id).
\item When a subscriber is no longer interested in a channel, the device can send an \emph{unsubscribe} request to the broker.
\end{enumerate}

The interactions made by the user will ultimately be translated into requests made to the broker from the subscriber's device application. Figure \ref{subsriber-broker-API} illustrates the types of communication between a broker and a subscriber's (``dharma1'') device. There are five main interactions that will occur between the broker and a subscriber to the \emph{EmergenciesNearMe} channel. The reason for using a pull-based architecture is to allow the device application to be intermittently connected and to use its own heuristics (including network connection, battery power, etc.,) to determine when and how to fetch results. In addition, the application may use heuristics to tell the broker how it wants to be able to get results (e.g., if I have been disconnected for a long time, just give me the newest result), which the broker can use to manage result caching.

The end subscriber does not need to be aware of the data cluster at all, or of how the channels and locations are being maintained. This separation of concerns is repeated in the broker-to-data-cluster interaction. 

\subsection{Broker Network}

The broker network is comprised of a scalable number of nodes, each designed to provide a one-to-many connection between the data cluster and the end user data subscribers. The broker network can capitalize on current Pub/Sub research for heuristics on subscriber distribution, subscription management, and result caching and delivery (as noted in Section \ref{requirements}). 
Such techniques are examined in \cite{uddinedge} and lie outside the scope of this paper. Below we focus on the communication layer between a broker and the data cluster. 

\begin{figure}[!ht]
  \centering
  \includegraphics[width=0.45\textwidth]{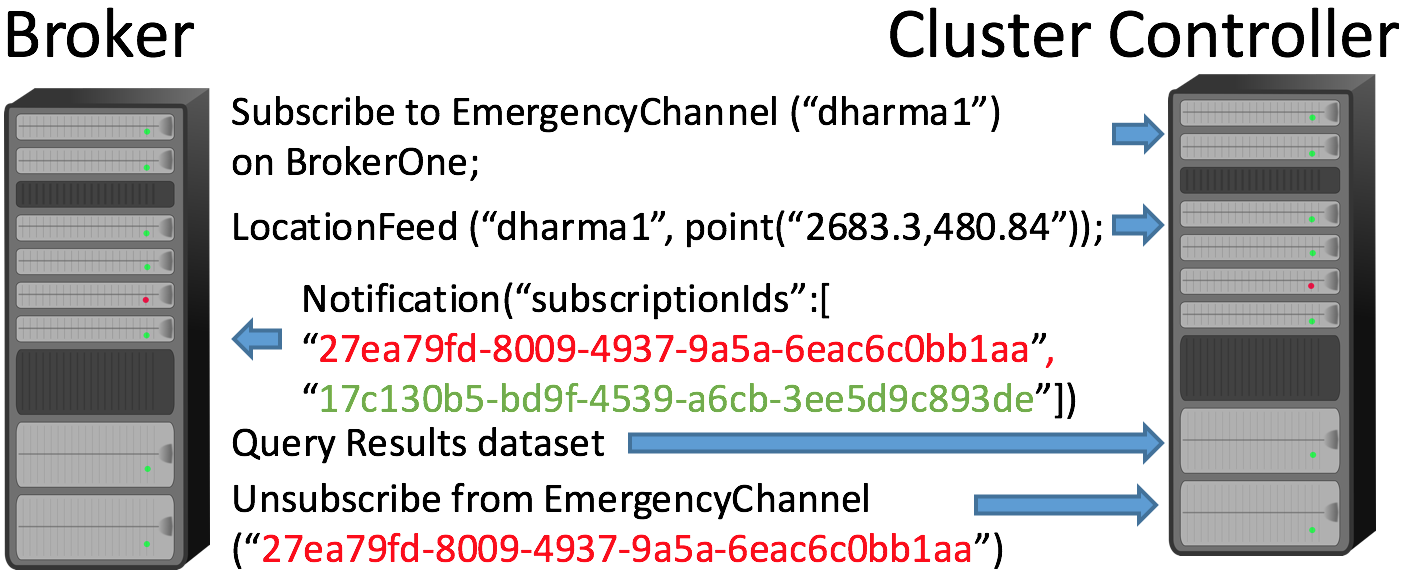}
    \caption{Communications between a broker and the data cluster}
  \label{broker-cluster-API}
\end{figure}

Requests made by an end data subscriber are handled (directly or indirectly) by requests made by the broker to the data cluster controller, using the following operations:

\begin{enumerate}
\item The broker sends a \emph{subscribe} request on behalf of a user (again, getting back a subscription id).
\item The broker sends newly reported user locations directly to the \emph{LocationFeed} data feed.
\item When the channel executes, the cluster sends a notification to the broker if there are new results, including the subscription ids for which there are new results.
\item The broker can run queries on the \emph{Results} dataset to retrieve results.
\item The broker can \emph{unsubscribe} on behalf of a user.
\end{enumerate}

It is up to the broker to determine when and how to query the \emph{Results} dataset based on the needs and availability of its users. In the simplest case, the broker can simply request results individually when they are requested by the user. 
The broker could fetch all of its users results every time it gets a notification, so these results would be cached for when users request them. A broker could also opt to only request and cache results for users who ask for results often, and hold off on other results \cite{uddinedge}. 

As an optimization, a broker could also share subscriptions among its users. For example, if there were a channel to find emergencies of a certain type in a certain city, there might be several users subscribing to tornadoes in San Francisco. Rather than creating a new subscription per user, the broker could share a single subscription on the cluster for these users but provide end user subscribers with individual ids to communicate their requests to the broker (which will use the shared subscription on the cluster).

\subsection{Data Cluster}
At the bottom of the Active Toolkit's software stack, the data cluster has the advantage of treating everything as ``just data''. Static and dynamic datasets, subscription data, result data, and user data all end up as scalable, distributed data that can capitalize on AsterixDB as a performant Big Data platform. 
This includes the ability to utilize AsterixDB's rich query capabilities for later ad hoc historical data analyses.
The abstract notions of \emph{subscribe} and \emph{unsubscribe} are translated at this level into simple cases of {\tt INSERT} and {\tt DELETE}. The application administrator can also build indexes on the datasets involved in the query, including the \emph{Subscriptions} and \emph{Results} datasets themselves. Temporal indexes can help with fetching recent data. The execution of the channel then becomes an optimized, scalable Big Data join.

\section{Pretending to be BAD} \label{glue}
To further showcase the advantages of building a BAD system, in this section we describe an alternative approach that attempts to address all the BAD desiderata (namely D1, D2 and D3 described in Section \ref{introduction}), by \textit{gluing} together existing off-the-shelf component systems (with minimum customization).  

To identify such components, consider again the emergency example application used throughout the paper. On the input side, we have the emergency reports and the user location updates, which are events that, depending on the application, can demonstrate high arrival rates and need to be efficiently ingested by the system. One could thus utilize an ingestion engine that is able to support such fast incoming data, e.g. Apache Kafka~\cite{kreps2011kafka}. 

On the output side, in order to deliver incoming data to subscribed users, one could utilize a notification delivery servicee like Amazon's Simple Notification Service (SNS)~\cite{amazonsns}. Amazon SNS enables notification delivery to a massive number of users and allows notification filtering based on a notification's content. 

Note that, in the BAD context, notifications' relevance to users do not consider just the content of the incoming data items, but also their relationships to other data (D1). 
Moreover, notifications sent to subscribers may need to be enriched (e.g., with shelter information) to provide actionable data (D2). Amazon SNS alone is thus not sufficient to support the D1 and D2 desiderata. The complete solution requires a system that also supports complex computations on large scale data in a timely manner; an example processing engine would be a streaming query processor like Apache Spark Structured Streaming~\cite{structured}.

Last but not least, ingested data (e.g., reports) can be used a posteriori for revealing useful insights, for example finding the frequencies and scopes of emergencies reported in the last year (see Figure \ref{historical_query}). To support such retrospective Big Data analytics (D3), we also need to persist the incoming data into a DBMS, e.g., MongoDB. Additionally, the large set of user subscriptions can also be persisted in this DBMS to remember them and efficiently support changes to them. 

\begin{figure}
\begin{minipage}[t]{0.5\textwidth}
{\small
\begin{sql}
  USE emergencyNotifications;

  SELECT report.Etype AS Emergency_Type, 
    count(report) AS Frequency, 
    avg(spatial_area(report.location)) AS Avg_Scope
  FROM Reports report
  WHERE report.timestamp > 
    current_datetime() - duration(``P1Y")
  GROUP BY report.Etype;
\end{sql}
\caption{An analysis of emergencies in the last year}
\label{historical_query}
}
\end{minipage}
\end{figure}

\begin{figure}
    \centering
    \includegraphics[width=.45\textwidth]{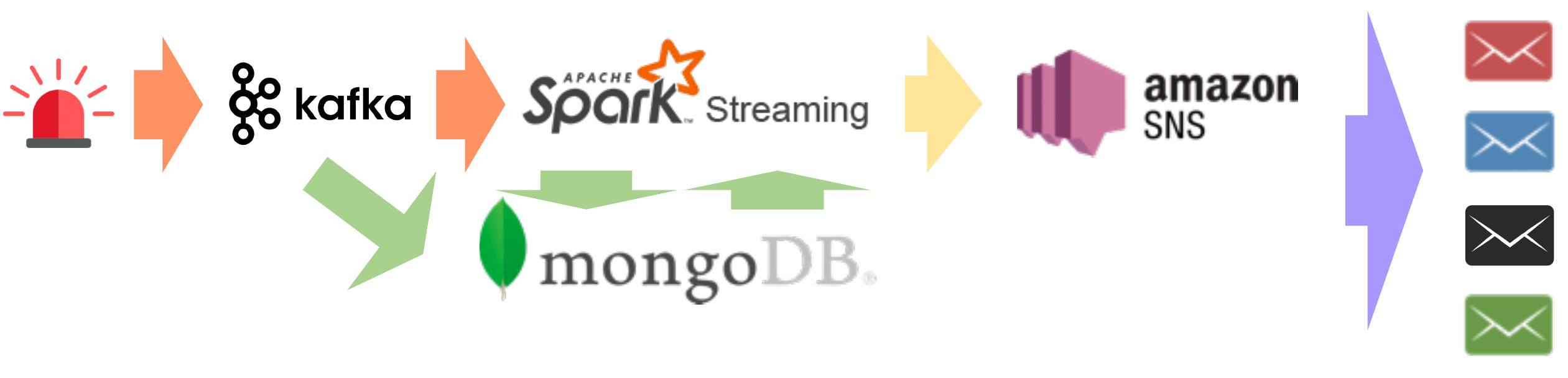}
    \caption{A `pretending to be BAD' example}
    \label{fig:not_bad}
\end{figure}

Figure~\ref{fig:not_bad} depicts an example of  \textit{gluing} all these four components together (namely the ingestion tool, the streaming engine, the data warehouse, and the notification service). While the figure handpicks a popular choice per component, there are many other choices one would have to consider and examine. For example, instead of AmazonSNS, one could consider Firebase Cloud Messaging, or Microsoft Notification Hubs, or even low-level customized web socket services.
Further, some choices (e.g. Kafka) may be used to substitute for multiple components (e.g. as an ingestion tool and a streaming engine), increasing the number of possible glueable combinations. 
Nevertheless, to build such a system, one would have to spend a significant amount of effort on configuring and orchestrating the different components. Gluing components together in this way involves disadvantages including the effort as well as potential runtime overhead and functional limitations. 

\noindent {\bf Limitations:} We can categorize the limitations of the gluing approach into three categories:

\noindent\textit{(1) Management Complexity.} Different systems have and utilize different data types and data representations in their runtimes. As data is transferred between different components, it may need to be transformed between different data types repeatedly. These translations may result in added overheads. For example, data items are ingested in Kafka as JSON strings. Spark then would parse these strings and cast the parsed result into rows in data frames, which then would be processed and persisted as documents in MongoDB. Eventually, data has to be transformed into JSON strings again for the AmazonSNS notifications. 
Further, a user would need to configure and deploy each component separately and then glue them together. Setting up the environment and maintaining it requires significant user effort and domain knowledge about its component. These additional management complexities are avoided when using the BAD platform's unified model.

\noindent\textit{(2) Limited Functionality.} The various components may not offer all of the needed functionality. Assume that we were to create Channel-like jobs in Spark. To create the notifications, one would need to run a spatial join between the report and user locations. However, Spark Structured Streaming does not support spatial joins over streams. 
This implies that a user would have to modify their existing queries to use equi-joins (thus limiting the application functionality). In contrast, in the BAD system, the user can take their existing queries and use them directly in channels.
As another limitation, Spark Streaming can only operate on a limited suffix of a data stream (due to memory limitations). As a result, if a user has not updated his/her location recently, such a location may not be available for the application.   

\noindent\textit{(3) Integration Difficulty.}
Given the presence of multiple independent components, data exchange between them in the glued system is inevitable and frequent. Although different vendors have provided connectors for bridging the gaps between them, users still would have to construct configuration files or even ``glue" programs for shipping data from one component to the other. Because of this, the system as a whole loses the possibility of \textit{optimizing} user queries across components. Data stored in MongoDB would first be pulled (via a full scan) into Spark for computation; Spark would then not utilize efficient data structures such as indexes to accelerate the data accesses in MongoDB. In the BAD system, users can create channels by just using SQL++ statements and they have no need to write lower-level programs. The channels can be optimized by the AsterixDB query optimizer to seek their most efficient query plans, and users can create indexes on datasets that BAD can then utilize to improve their runtime performance.

The above discussion summarizes why the BAD approach is unique and is not directly comparable with any one alternative platform -- the only functional alternative is to construct a multi-system tangle. 
We further note that if one were to avoid dealing with the glue issues among multiple components, picking just one component and heavily customizing it to meet the remaining BAD requirements, the task would be challenging (or even impossible) since each component provides only a subset of the required desiderata.
No one system has the persistence, query power, and declarative-ness of BAD. For example, using only Spark Streaming, one would have to customize ingestion and result delivery. Spark could persist data in HDFS, but without database guarantees (updates, consistency, concurrency). Similarly, Kafka is not designed for storing data and has limited querying capabilities. Amazon SNS is a data routing service without complex computation capabilities or storage.

Loosely speaking, the end goal of BAD is to reduce the effort required to build big active data applications in a manner not unlike the way that the onset of relational databases and SQL reduced the effort required to build passive business applications -- it should be possible to build applications declaratively, with a minimum of programming effort.

\section{Experimental Evaluation} \label{experiments}
We now proceed to examine how our initial implementation of a BAD system performs and scales. 
Our objective in this context is simply to take a first look at the performance characteristics of the BAD approach itself. 
(We leave the possibility of comparing BAD to a wired-together glue competitor to future work.)

The separation of concerns between the data subscribers, the brokers, and the data cluster allows us to separate their performance evaluations. For example, one could look at the end-to-end performance experienced by users, the caching and user distribution performance of brokers, or the data cluster itself. As this paper is focused on the techniques and research of the BAD data cluster, our experiments here focus on this layer. Specifically, we look at three performance aspects of the data cluster: ingesting scalable data, processing channel jobs for a scalable number of users, and delivery of results to the brokers. We see this as a critical factor of the performance overall, as all other aspects of performance will depend on it. Since users are just ``data'' from the data cluster perspective, we can essentially remove the end users from the performance picture and treat the brokers as the ``clients'' of the BAD data cluster. 

\subsection{Experimental Setup}
For our experiments, we used the example application that has been discussed throughout this paper, including its data model, datasets, feeds, and the repetitive EmergenciesNearMe channel with period of 10 seconds (Figure \ref{channel_ddl}). It is important to note that the choice of 10 seconds is not important in and of itself. Rather than fixing the window size and varying the arrival rates of data, we could fix the arrival rates and vary the window size to do a similar evaluation. The definitions of data types, data feeds, channel, and broker were shown in Figures \ref{aql-example}, \ref{feeds-ddl}, \ref{connfeeds-ddl}, \ref{function}, and \ref{channel_ddl} respectively.

In order to model the realistic movement of people in the real world, we used the the Opportunistic Network Environment (ONE) simulator \cite{one:website,keranen-theone,desta2013evaluating} to simulate users' movement in our experiments. The ONE simulator allows for maps to be built representing real cities, including map graphs of pedestrian paths, roads, tram routes, etc., and for creating classes of ``hosts'' that represents cars, pedestrians, and commuters by providing movement models and graphs for each class, thereby simulating a realistic flow of human movement within that city. The publicly available ONE simulator comes with a pre-built simulation of the city Helsinki, including the actual roads and metro routes. We used this city in our following experiments.

Due to the complexity of modeling massive user movement directly in the ONE simulator, we let the ONE simulator generate data for small groups of users, and then we merged them together for our experiments depending on the number of users that we needed. In each group, there were 30 users, which includes 10 pedestrians, five cars, and 15 metro commuters. The metro commuters were evenly assigned to three different metro routes, five commuters for each route. The moving users reported their location every 10 seconds in our simulated world where there were emergencies happening at the same time. Our goal was to see how many users we can support with our system in the given emergencies-near-me application.

\begin{figure}[!ht]
  \centering
  \includegraphics[width=0.45\textwidth]{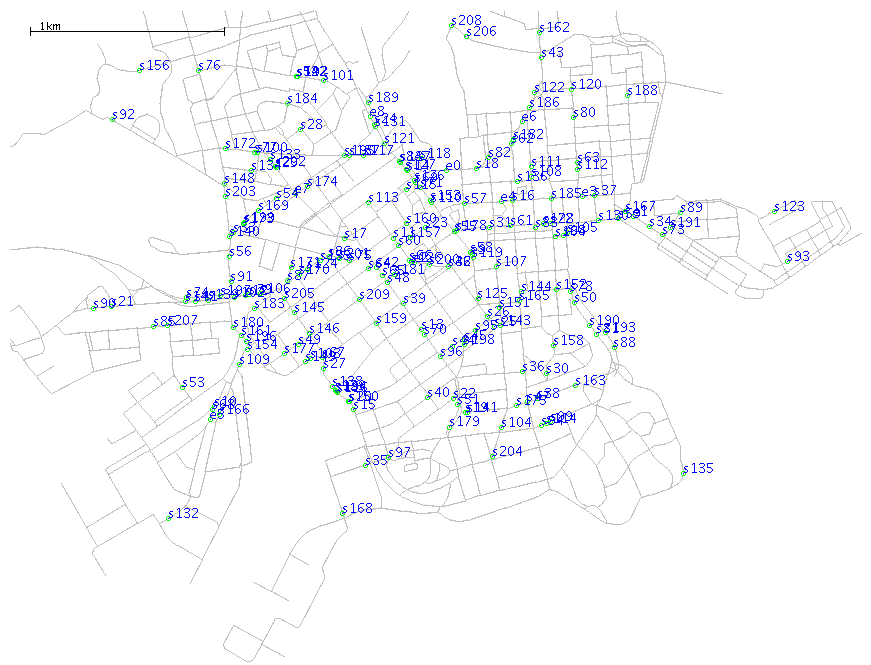}
    \caption{Distribution of 200 shelters in Helsinki}
  \label{200Shelters}
\end{figure}

The emergencies were generated using the ONE simulator as well. We created a group of ten emergency creators and allowed them to traverse rapidly and randomly around the map. There were four potential emergency types: floods (with a radius of about 1/8th of the city and a probability of 50\%), fires (with a radius of about 1/16th of the city and a probability of 30\%), storms (with a radius of about 1/4th of the city and a probability of 10\%), and car crashes (with a radius of about 1/100th of the city and a probability of 10\%). Also, we randomly generated a set of statically located shelters in Helsinki. A distribution of the 200 shelters is shown in Figure \ref{200Shelters}.

With this emergency-heavy scenario setup, which one might characterize as ``Hell''-sinki, we ran the ONE simulator with the described configurations, exported the data, and converted them into AsterixDB data model (ADM) files, so we could repeatedly replay the emergencies-near-me scenario by reloading the generated data. In each experiment, we loaded the shelter data into the \emph{Shelters} dataset and then fed the users' locations and emergencies through the ``LocationFeed' and ``ReportFeed' into the \emph{UserLocations} and \emph{Reports} datasets respectively.

\subsection{Data}
To start with a non-empty state, we pre-loaded the \emph{Reports} dataset with a history of over 2 million reports. Since we created an index on the ``report\_time'' attribute, the size of the report dataset would not harm the overall performance of the channel execution, but we pre-loaded so as to show this held true. The \emph{Shelters} dataset was pre-loaded with 200 shelters scattered across Hellsinki.

In our experimental application, there were two parameters that could be changed: the arrival rates of new reports, and the number of users. We assumed all of our users were subscribed to the channels, thus the subscriptions dataset had the same size as the \emph{UsersLocations} dataset. A new emergency report coming into the system through the ``ReportFeed'' was automatically assigned a UUID as its primary key and then stored in the \emph{Reports} dataset. A user sent his/her location update every 10 seconds through the ``LocationFeed'', and this location was then upserted into the \emph{UsersLocations} dataset. We only kept the latest location of each user in our system.

\begin{figure}[!ht]
  \centering
  \includegraphics[width=0.45\textwidth]{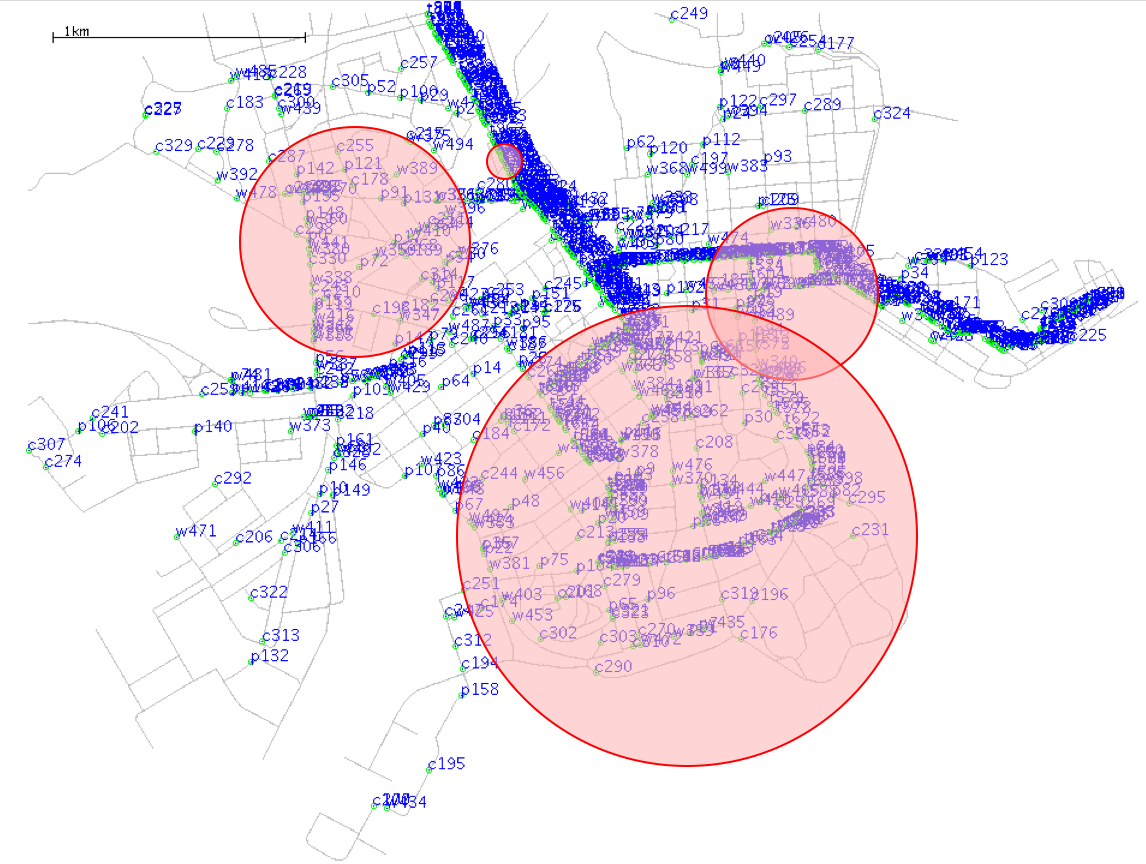}
    \caption{A BAD moment in ``Hellsinki''}
  \label{1000UsersWithEmergencies}
\end{figure}

Figure \ref{1000UsersWithEmergencies} shows a snapshot in time of ``Hellsinki''. Here we show the locations of 1000 users along with the locations of four emergencies that occurred. The dense areas of users are commuters on the city's tram routes.

\subsection{Experimental Setup} \label{timeCompare}
In order to show the advantages of using the BAD system, we designed another approach that supports the same scenario using only the passive AsterixDB. In this approach, the broker would need to send explicit requests to AsterixDB on behalf of every user. We created a ``polling'' program that reads users' location updates and issues the query shown in Figure~\ref{polling-query} to obtain the the nearby emergencies and shelters within the same time interval as the repetitive channel (10 seconds). The ``polling program'' maintained a queue to receive the incoming users' location updates, and there are one or more ``poller'' threads that take the location updates from the queue and query AsterixDB for the requested information. We will further investigate how the number of ``poller'' threads affects the performance later in Figure \ref{ThreadComparisonGraph}. For clarity we will refer to BAD AsterixDB as the \emph{active} mode and to vanilla AsterixDB as the \emph{passive} mode.

\begin{figure}
\begin{minipage}[t]{0.5\textwidth}
{\small
\begin{sql}
USE emergencyNotifications;

SELECT r, shelters
FROM Reports r,
  (SELECT s.location FROM Shelters s 
    WHERE spatial_intersect(s.location,
    circle(``2437.3,1330.5 100.0"))) shelters
WHERE r.timestamp > current_datetime() - 
  day_time_duration("PT10S")
  AND spatial_intersect(r.location, point(``2437.3,1330.5"));

\end{sql}
\caption{The polling query for a single user (at location ``2437.3,1330.5'')}
\label{polling-query}
}
\end{minipage}
\end{figure}

In the active (i.e., BAD) mode, we loaded the \emph{Shelters} and \emph{Reports} datasets with the shelters data and the prepared 2 million history reports first. Then, we created the EmergencyNearMe channel and loaded the subscription data. Lastly, we started feeding data into the \emph{UserLocations} and \emph{Reports} datasets with a specified number of users and report rates. Our experiments measured how well the channel was able to scale in terms of the number of supportable users, i.e., how many users and how much data it could handle within the desired inter-results period of 10 seconds. After that, the system would be in a ``overloaded'' situation since it would fail to operate within the specified interval of the channel. We ran the experiment multiple times to search for the maximum supportable users given a certain rate of reports.

To instrument the active experiment, we recorded the time ($t_e^i$) of the channel (`Channel Execution Time', i.e., the time to produce results), and the time ($t_f^i$) that the broker spent on fetching the new results on behalf of users after receiving a notification (`Result Fetching Time') in all $n$ invocations throughout the channel lifetime, where $0 \le i \le n$. Notice that both the Channel Execution Time and the Result Fetching Time need to be within the 10-second period for a channel to work in a stable state. The criterion that the channel can support a given number users and a certain report rate is $\max_{i = 1}^{n}(\max(t_e^i, t_f^i)) < 10$. 

In the passive (i.e., user polling) mode, we initialized the \emph{Shelters} and \emph{Reports} datasets in the same way using the prepared shelter and reports data and then started feeding only the \emph{Reports} dataset. After that, we started the poller program that reads users' location updates and polls the requested emergencies and shelters for each user location one-by-one using the query in Figure~\ref{polling-query}. For a given number of users, the poller program has to be able to consume all location updates from all users in each 10-second interval, otherwise it too would be in an overloaded state. We measured how many users the poller program can serve given a certain report rate in the 10-second window.

\subsection{Hardware Configurations}
Our experiments were conducted on a six node cluster consists of Inter NUC (BOXNUC5I7RYH). Each node had an i7-5557U CPU processor (4 cores per machine), 16 Gigabytes of RAM, and a 1 TB hard drive. The nodes were connected with a Gigabit Ethernet switch. For the active mode, we deployed the BAD system on four nodes where one node hosted the Cluster Controller (CC), which accepts and compiles user queries and schedules query evaluation over the cluster. The other three nodes hosted Node Controllers (NC), which receive complied query plans from the CC, compute according to the plan, and store the data. Each NC contained one data partition. The broker and the data feeding programs which sent reports and location updates ran on the other two nodes separately. For the passive mode, we deployed the AsterixDB cluster using the same configuration as BAD, and the poller and reports feeding programs ran on the other two nodes.

\subsection{Feeds \emph{vs.} Manual Inserts}
\begin{figure}[!ht]
  \centering
  \includegraphics[width=0.45\textwidth]{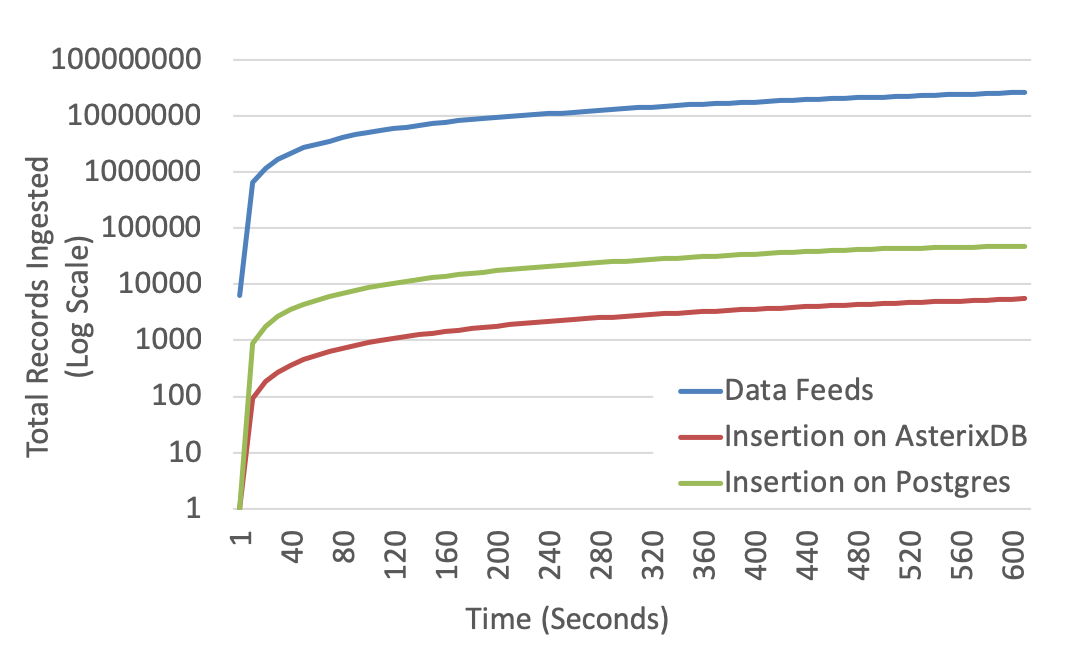}
    \caption{Feeds vs. Manual Inserts}
  \label{FeedPerformance}
\end{figure}

We used active data feeds for ingestion of both \emph{Reports} and \emph{UserLocations}. Data feeds are an important contributer of data for scalable channels, as they provide the mechanism used by the BAD platform to ingest data at scale. To illustrate the advantages of using data feeds for rapid data ingestion, we compared the ingestion performances of using data feeds versus issuing insert statements. We also evaluated against the insert performance of Postgres as another baseline. For data feeds, we set up an external program that fed data continuously. For passive AsterixDB and Postgres, we also set up an external program that inserted new data by repeatedly issuing insert statements. We measured the number of ingested data records during a 10 minutes experiment to show their performance differences. The experimental resutls are plotted in Figure~\ref{FeedPerformance}. Note that the number of ingested records is on logarithmic scale.

As we can see, the data feeds maintained very high ingestion performance during the experiment. Both passive AsterixDB and Postgres, issuing insert statements, had much lower ingestion performance. Data received by data feeds is parsed and fed to the ingestion pipeline which can then be persisted into the storage system directly. This allowed the incoming data to be efficiently consumed by the system. For the passive insertion case on both AsterixDB and Postgres, however, each insert statement has to be handled by the query compiler separately and executed as an independent job. This increased the per-record insertion cost, which led to their lower performance.

\subsection{Channels \emph{vs.} Polling}
\label{bad_expr}
In real scenarios, the published reports may have different degrees of intersection with the subscribed users. This intersection rate will affect the overall performance of both the BAD and polling approaches (channel execution time and result fetching time in the active mode, and query evaluation time in the passive mode). In order to explore how different intersection rates of reports and subscribed users affected performance, we designed four scenarios and compared the performance of each under the active (BAD) mode and passive (polling) modes respectively. The scenarios are as follows:

\begin{itemize}
\item Case 1: A large percentage of subscribers intersect with emergencies, and a large percentage of the emergencies intersect with subscribers.
\item Case 2: A large percentage of subscribers intersect with emergencies, but a small percentage of the emergencies intersect with those subscribers.
\item Case 3: A small percentage of subscribers intersect with emergencies, while a large percentage of the emergencies intersect with those subscribers.
\item Case 4: A small percentage of subscribers intersect with emergencies, and a small percentage of the emergencies intersect with those subscribers.
\end{itemize}

In order to demonstrate these scenarios, we created two additional cities outside of Hellsinki: Tartarusinki, where lots of emergencies happen but (fortunately) no one resides, and Elysinki, where lots of people reside but (also fortunately) no emergencies ever happen. The added cities each also have 200 shelters. In all cases, we did our breaking-point analysis, which showed how many subscribers the system can serve within the 10-second window for a given arrival rate of reports.

\subsubsection{Case 1 - ``Hellsinki'' Alone}
\label{sec:case1}
For Case 1, we assumed that Helsinki was the only city (as in Figure \ref{1000UsersWithEmergencies}). All emergencies took place there, and all subscribers resided there. As the number of incoming emergencies per second increased, Helsinki gradually became more and more of an apocalyptic ``Hell''-sinki.

\begin{figure}[!ht]
  \centering
  \includegraphics[width=0.45\textwidth]{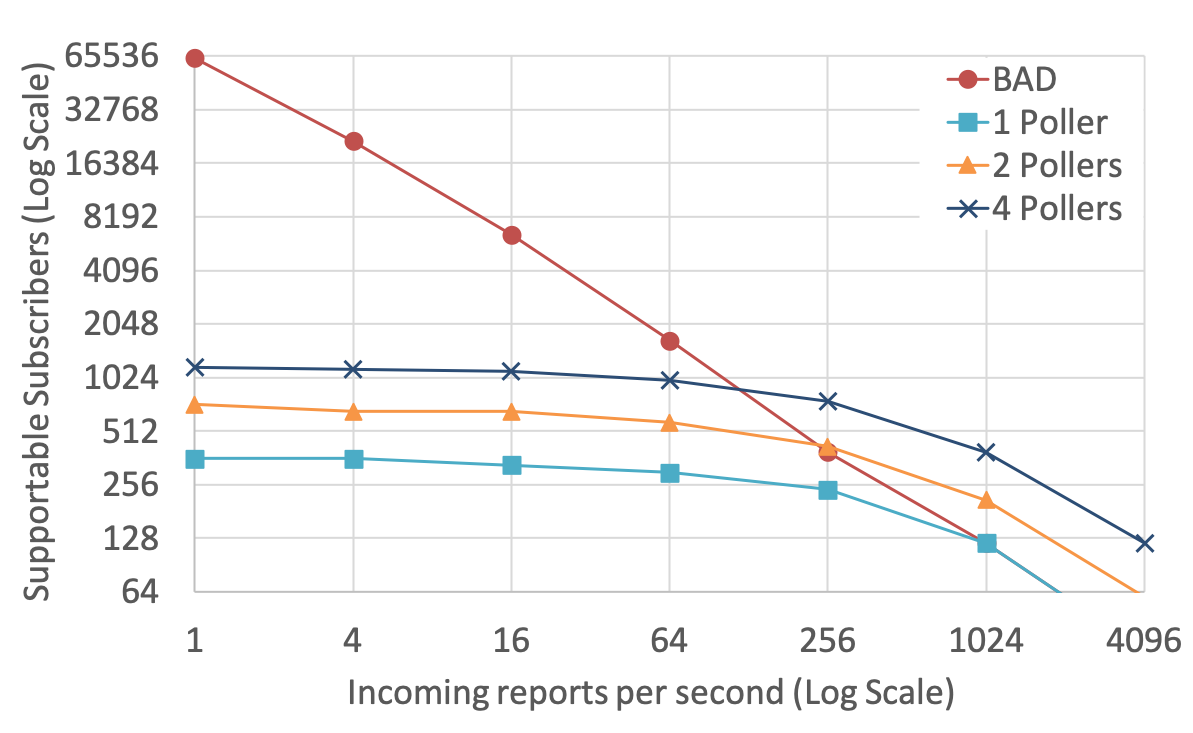}
    \caption{Case 1 - ``Hellsinki'' Alone}
  \label{HellsinkiGraph}
\end{figure}

We present the experimental results in Figure~\ref{HellsinkiGraph}. The x-axis represents the rate of reports. The y-axis shows the maximum number of subscribers that can be served while staying within the 10-second delivery deadline. Note that both the x and y axes are on a logarithmic scale due to the large performance differences observed.

In the passive mode, the polling performance is affected by not only the execution cost of computing the results but also the compilation cost in processing the polling queries. When AsterixDB receives a polling query, it compiles and optimizes that query into a job specification on the CC node and sends that specification to the NC nodes for execution. When the report arrival rate is low, the pollers' performance is mainly bounded by the available resources on the CC node for compiling the polling queries, so the pollers' performance is relatively stable on the left side of the graph (i.e., for smaller report rates). As the report arrival rate increases, the high execution cost of each polling query causes fewer queries to be completed during the given time window, so the execution cost becomes the main performance impactor. As a result, the pollers' performance drops when the report arrival rate is very high. Increasing the number of pollers improves the performance (but within a limit); we will further discuss this in Section~\ref{discussion}. 

The active mode starts with a drastically higher number of supportable subscribers that gradually decreases as the rate of reports increases. The decrease is because the computational load for the channel query execution grows when there are more reports being generated during its execution window (10 seconds in this case). 
The active mode outperforms or (at worst) matches the one poller passive mode. It is strikingly better for lower incoming report rates, where it outperforms the one poller passive mode by supporting up to two orders of magnitude more subscribers.
The passive mode with multiple pollers outperforms the active mode only for much higher rates (many hundreds of incoming emergency reports/sec). We will further analyze the performance of the BAD system versus the passive mode in Section~\ref{discussion}.

\subsubsection{Case 2 - Hellsinki and Tartarusinki}
\begin{figure}[!ht]
  \centering
  \includegraphics[width=0.45\textwidth]{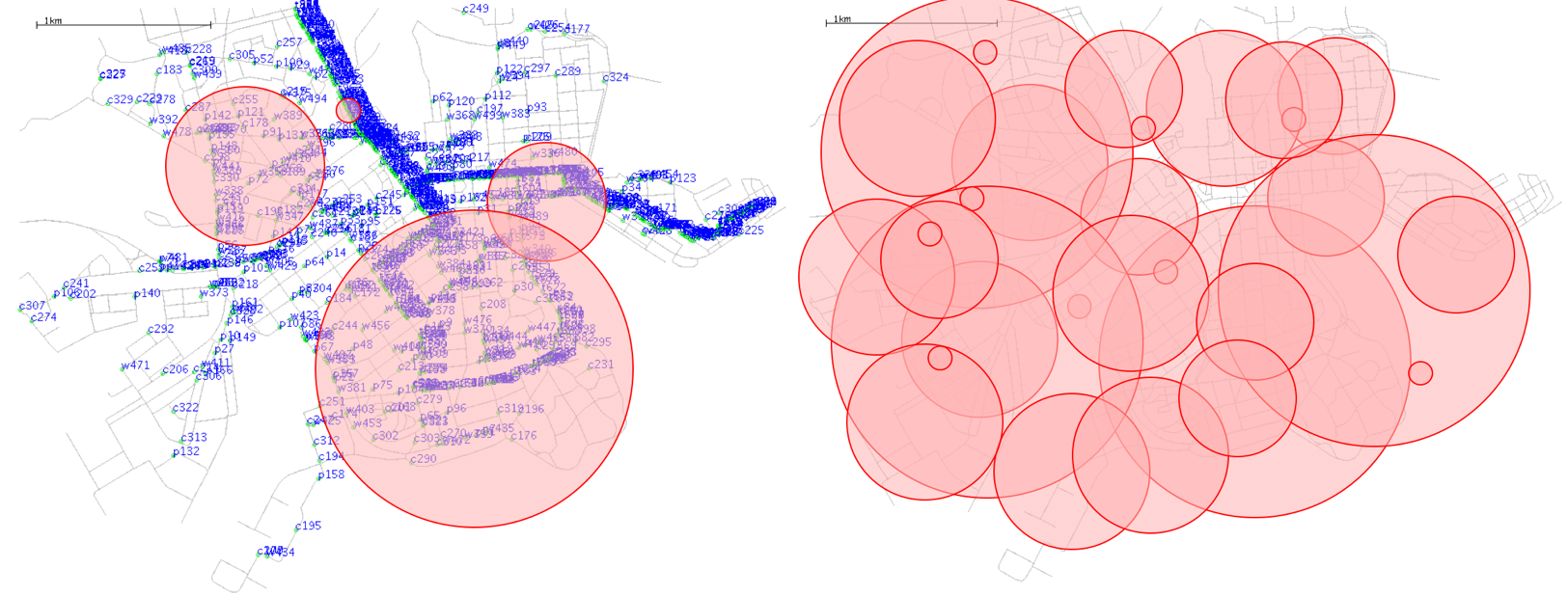}
    \caption{Hellsinki (Left) next to Tartarusinki (Right)}
  \label{TartarusinkiContingent}
\end{figure}

If we add Tartarusinki to our map and have 90\% of all of the emergencies occur there, all of our subscribers will potentially receive notifications, but most of the emergencies that occur ``worldwide'' will not contribute to those notifications. Figure \ref{TartarusinkiContingent} shows this scenario. 

\begin{figure}[!ht]
  \centering
  \includegraphics[width=0.45\textwidth]{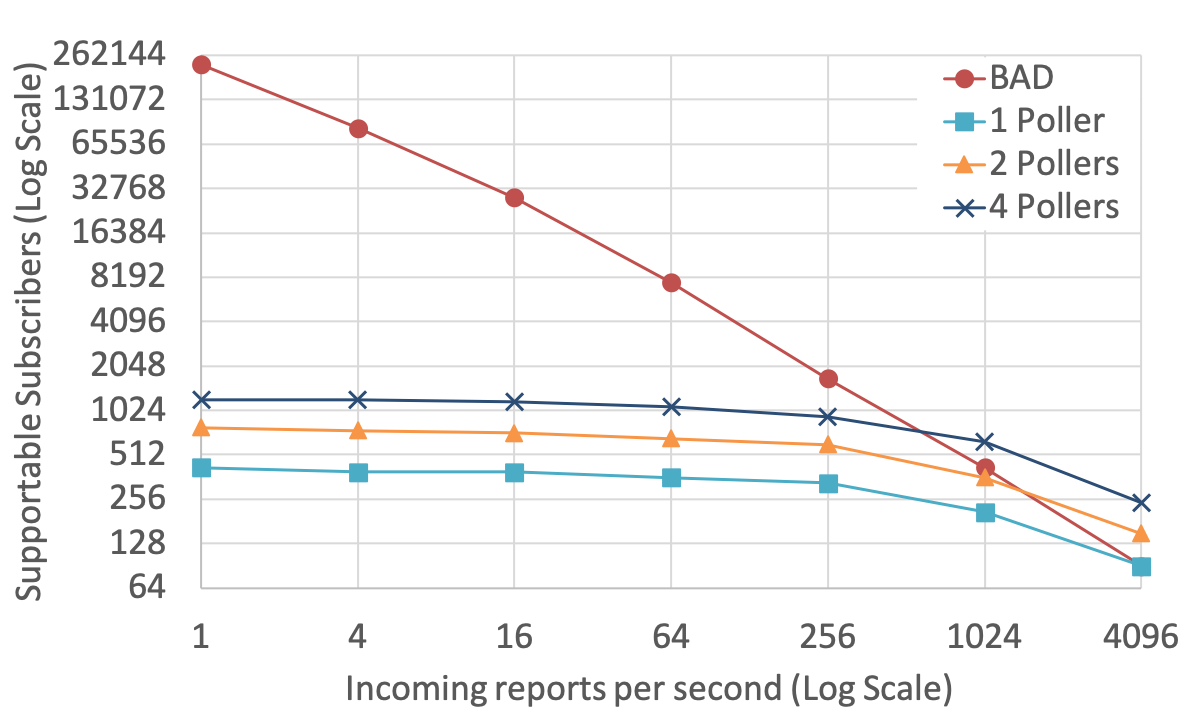}
    \caption{Case 2 - Hellsinki and Tartarusinki}
  \label{case2}
\end{figure}

As we can see from the results shown in Figure~\ref{case2}, the active mode starts with a much higher number of supportable subscribers compared with Case 1. This is because in Case 2 there are many less notifications generated given the same number of subscribers and rate of reports. In the passive mode, the primary bottleneck is still the query compilation cost on the CC node. Thus, the passive mode's performance in Case 2 is similar to that in Case 1, and it remains relatively stable for smaller report rates. As the report rate increases, the passive performance starts to decrease due to a larger workload introduced by more reports in the channel execution window.
Note that in Case 2, the BAD system's performance superiority with respect to the passive mode with multiple pollers extends further to the right; as before the active mode outperforms the one poller passive case, except for very high report rates where the two approaches behave similarly.

\subsubsection{Case 3 - Hellsinki and Elysinki}
\label{sec:case3}
\begin{figure}[!ht]
  \centering
  \includegraphics[width=0.45\textwidth]{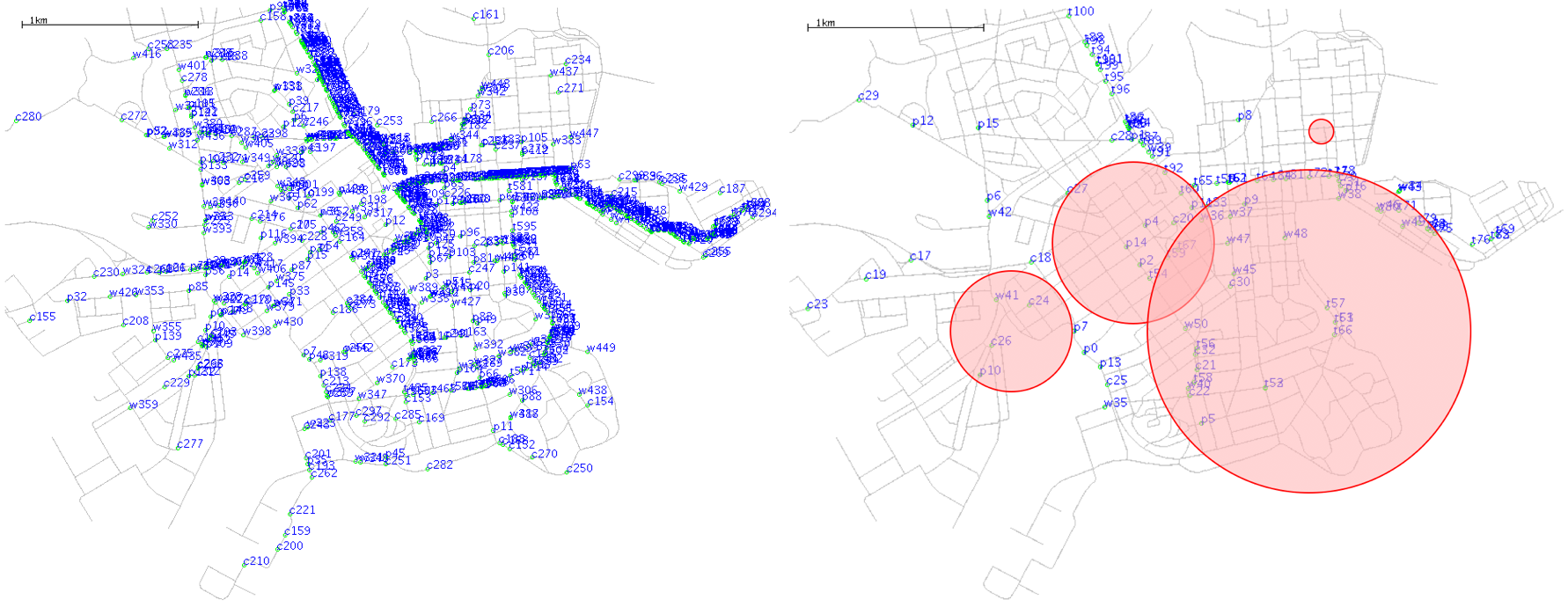}
    \caption{Elysinki (Left) next to Hellsinki (Right)}
  \label{ElysinkiContingent}
\end{figure}

If instead we add Elysinki as the second city, all of the emergencies will be in Hellsinki. Since only 10\% of the subscribers are in Hellsinki now, most subscribers will not receive emergency notifications. Figure \ref{ElysinkiContingent} shows this scenario. 

\begin{figure}[!ht]
  \centering
  \includegraphics[width=0.45\textwidth]{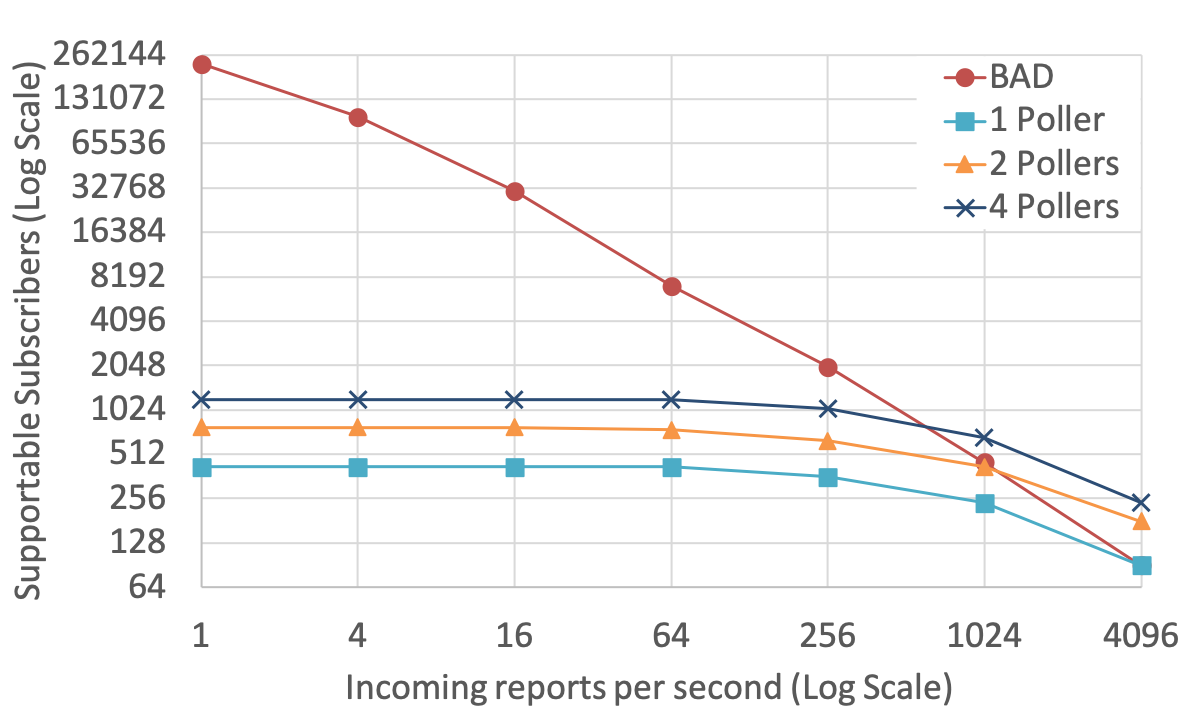}
    \caption{Case 3 - Hellsinki and Elysinki}
  \label{case3}
\end{figure}

Figure \ref{case3} shows the related performance results. Although the subscribers outside of Hellsinki are involved in the channel computation, there are no results produced for them. Thus, the active mode starts with a much higher number of supportable subscribers compared with Case 1. Note that the general performance trend of Case 3 resembles the performance in Case 2. This is expected as we reduced 90\% of the ``effective'' reports in Case 2 by moving them to Tartarusinki, and we reduced 90\% of the ``effective'' users in Case 3 by moving them to Elysinki. The number of produced results in both cases are roughly the same, which leads to similar performance. This is also supported by the results in Section \ref{whereTimeGoes} which presents the channel execution times and result sizes for all cases.
As before, the passive mode performance remains stable when the rate of reports is low and starts to drop when it further increases.

\subsubsection{Case 4 - All three cities}
\begin{figure}[!ht]
  \centering
  \includegraphics[width=0.45\textwidth]{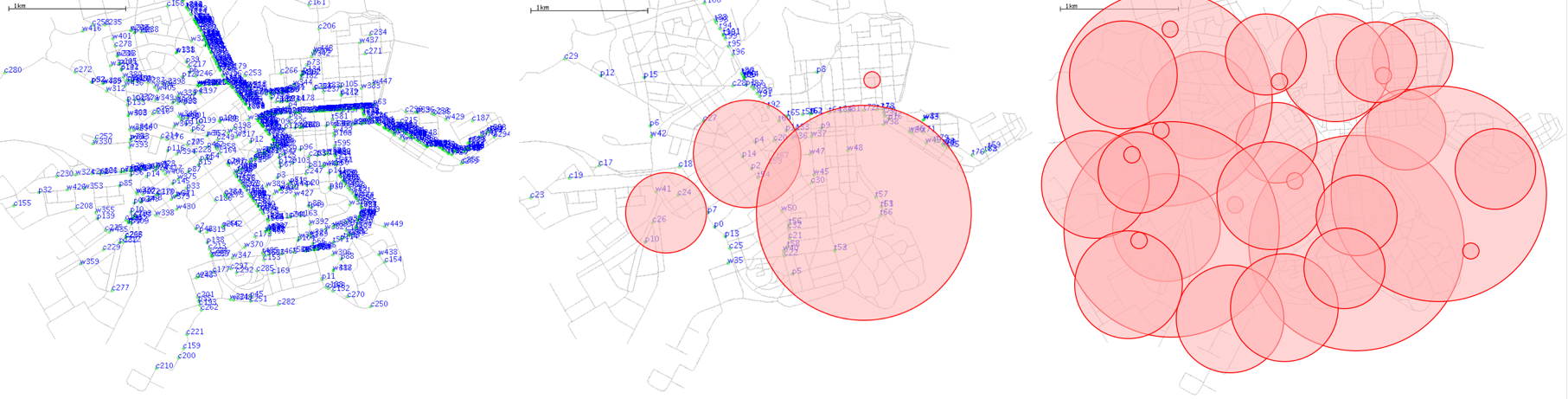}
    \caption{Elysinki(Left), Hellsinki(Center), and Tartarusinki(Right)}
  \label{World}
\end{figure}

Lastly, we simulate a world where 90\% of the emergencies occur in Tartarusinki while 90\% of the subscribers reside in Elysinki, and an unlucky few remain in Hellsinki. In this case emergency notifications will exist for only 10\% of both emergencies and subscribers. Figure \ref{World} depicts this scenario.

\begin{figure}[!ht]
  \centering
  \includegraphics[width=0.45\textwidth]{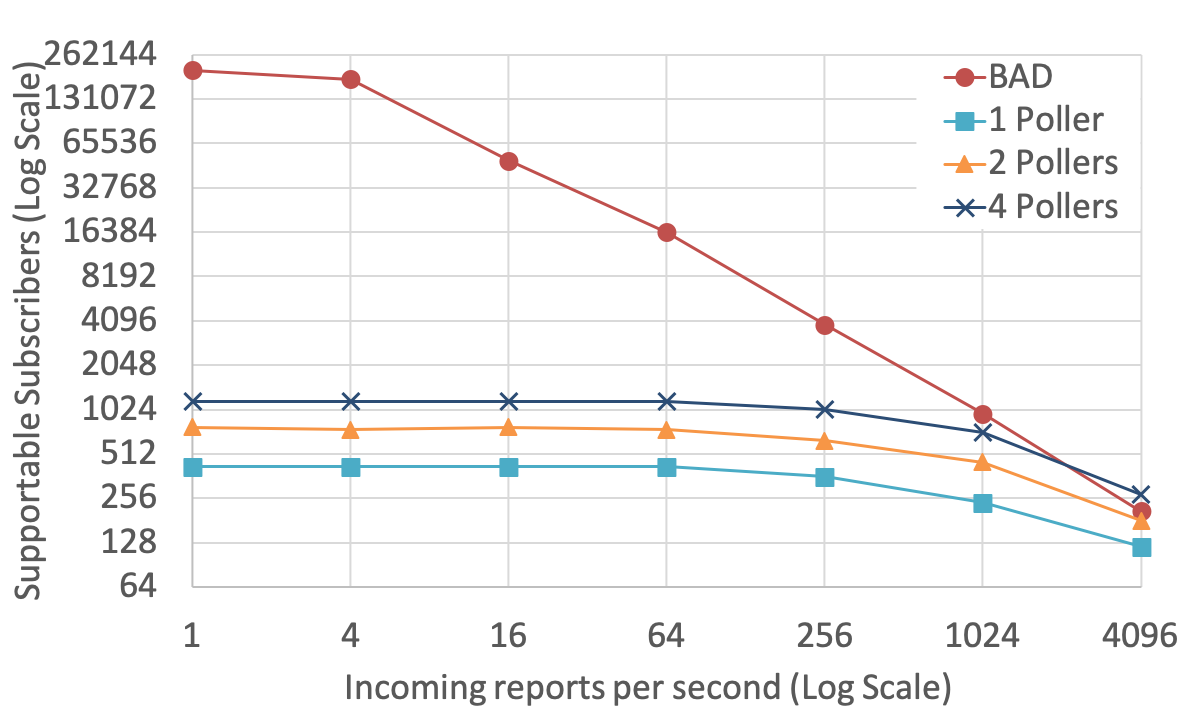}
    \caption{Case 4 - All three cities}
  \label{case4}
\end{figure}

As shown in Figure~\ref{case4}, the active mode outperforms the one and two pollers passive mode even when we scale the disaster rate to thousands of emergencies per second. Different from Cases 2 and 3, the performance of the active mode does not drop as much as we increase the rate of reports from one to four. This is because of the low intersection rate of the ``effective'' reports and users, which causes very few results to be generated when the report rate is relatively low. After we further increase the report rate, performance starts to drop as before.

The performance of the passive mode is similar to the previous cases. It remains stable for smaller report rates and starts to drop at very high report rates. 
When considering all four scenarios, the main difference in the passive mode is that for Case 1, the performance deterioration starts at much lower report rates (around 64 reports/sec). This is because for Case 1, even a relatively small report rate produces large computational workload (due to the large number of intersections between reports and users).

\subsubsection{Discussion}
\label{discussion}
\begin{figure}[!ht]
  \centering
  \includegraphics[width=0.45\textwidth]{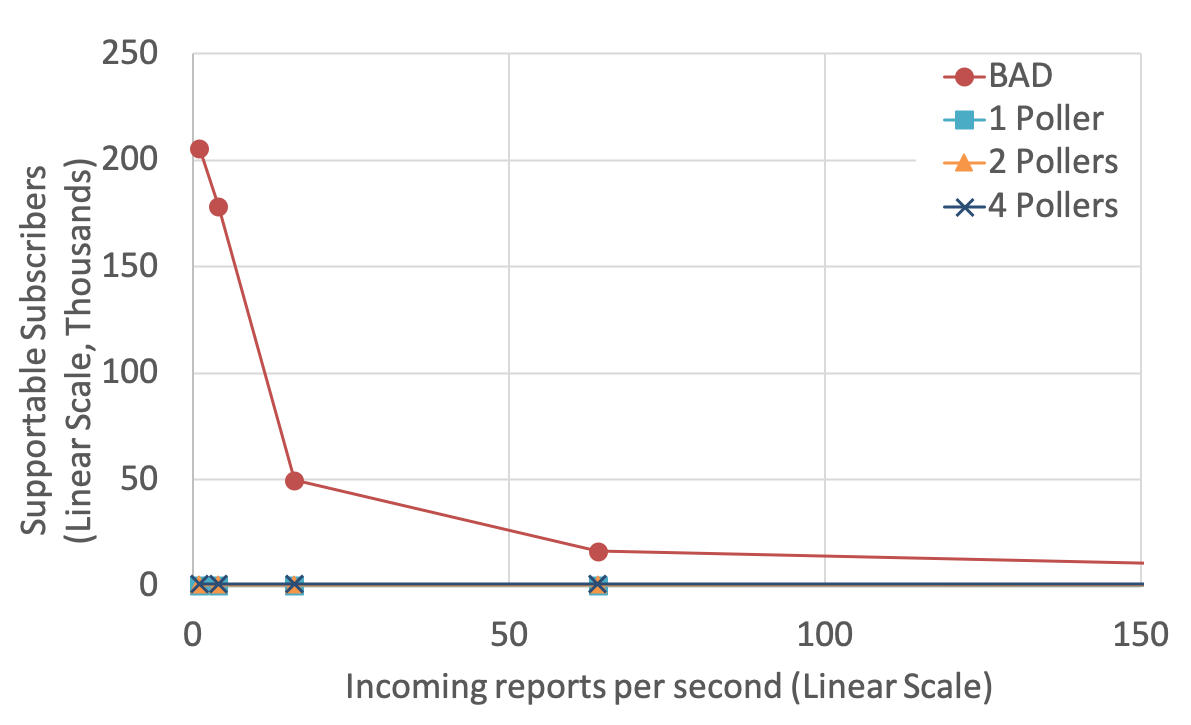}
    \caption{Case 4 on a linear scale}
  \label{case4linear}
\end{figure}

\begin{figure}[!ht]
  \centering
  \includegraphics[width=0.45\textwidth]{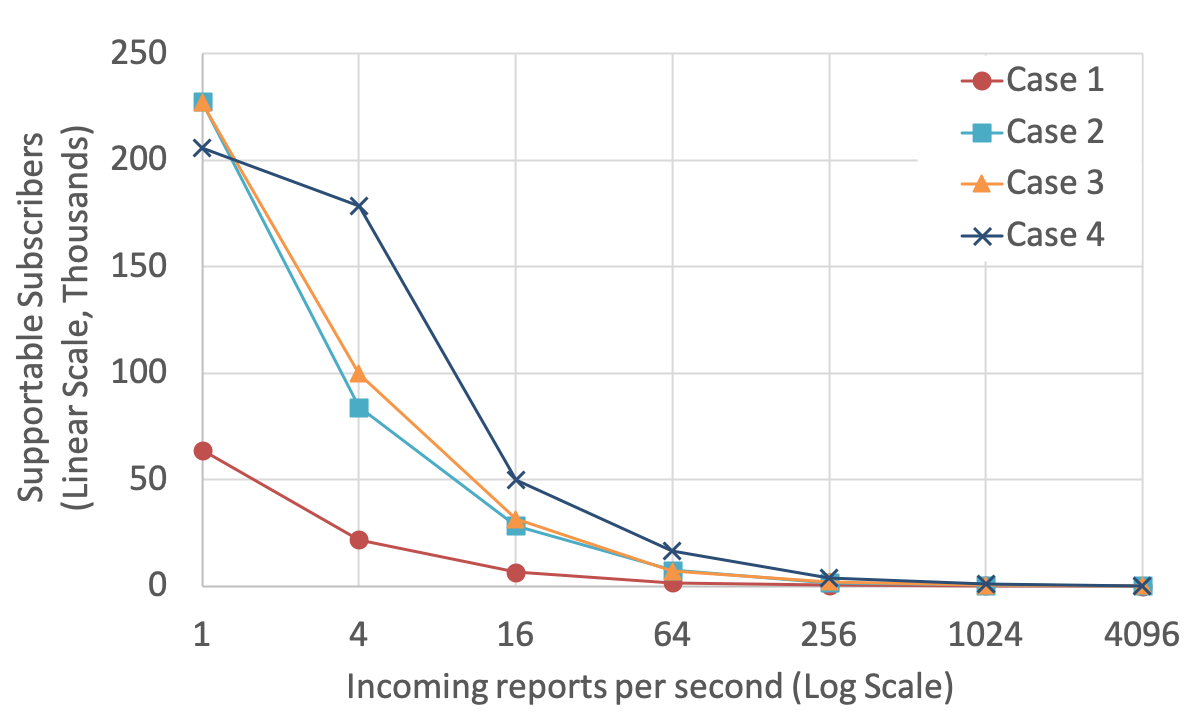}
    \caption{Active performance in the four cases}
  \label{casecomparechannels}
\end{figure}

The previous graphs all used logarithmic scales. Figure~\ref{case4linear} shows a portion of the performance graph from Case 4 (Figure~\ref{case4}) using a linear scale to better convey the large performance difference between the active and passive modes. The performance of the active mode for all four cases is summarized for comparative purposes in Figure~\ref{casecomparechannels}. Case 1 represents the most `apocalyptic' scenario where all events and all subscribers have the potential to intersect. Thus, the number of supported subscribers is the lowest. Cases 2 and 3 each have only a fraction of Case 1's ``effective'' data (10\% reports in Case 2 and 10\% users in Case 3), so their performance is better than in Case 1, and the performance of Cases 2 and 3 is very similar. 
In Case 4, there are three cities, which increased the number of shelters (we used 200 shelters per city, so Cases 2 and 3 had 400 shelters each, while Case 4 had 600). Thus, the starting point of Case 4's performance (1 report per second) is slightly lower than Cases 2 and 3 due to the additional computational cost from joining with more shelters. 
The performance of Case 4 drops slower than Case 2 and Case 3 while the report rate increases, since in Case 4 there are fewer results generated (there are fewer intersections as most users and most reports are moved to different cities).

For low to moderate report rates, the BAD system is much better than the passive mode in all four cases. This shows the advantage of batch processing and deployed jobs used in the BAD system, as opposed to repeatedly issuing the polling queries using poller programs in the passive mode. When the report rate becomes very high, the performance of both the BAD and passive mode start to decline due to the increased workload. In particular, the gap between the BAD system and the passive mode with one poller narrows and their performance eventually converges. This is because the BAD executes one channel query in each time window, and the one poller program makes one request for a user at a time. In both cases, there is only one query/job being executed concurrently. The resources allocated for evaluating the channel/polling query in both cases (the BAD and one poller) are roughly the same.

\begin{figure}[!ht]
  \centering
  \includegraphics[width=0.45\textwidth]{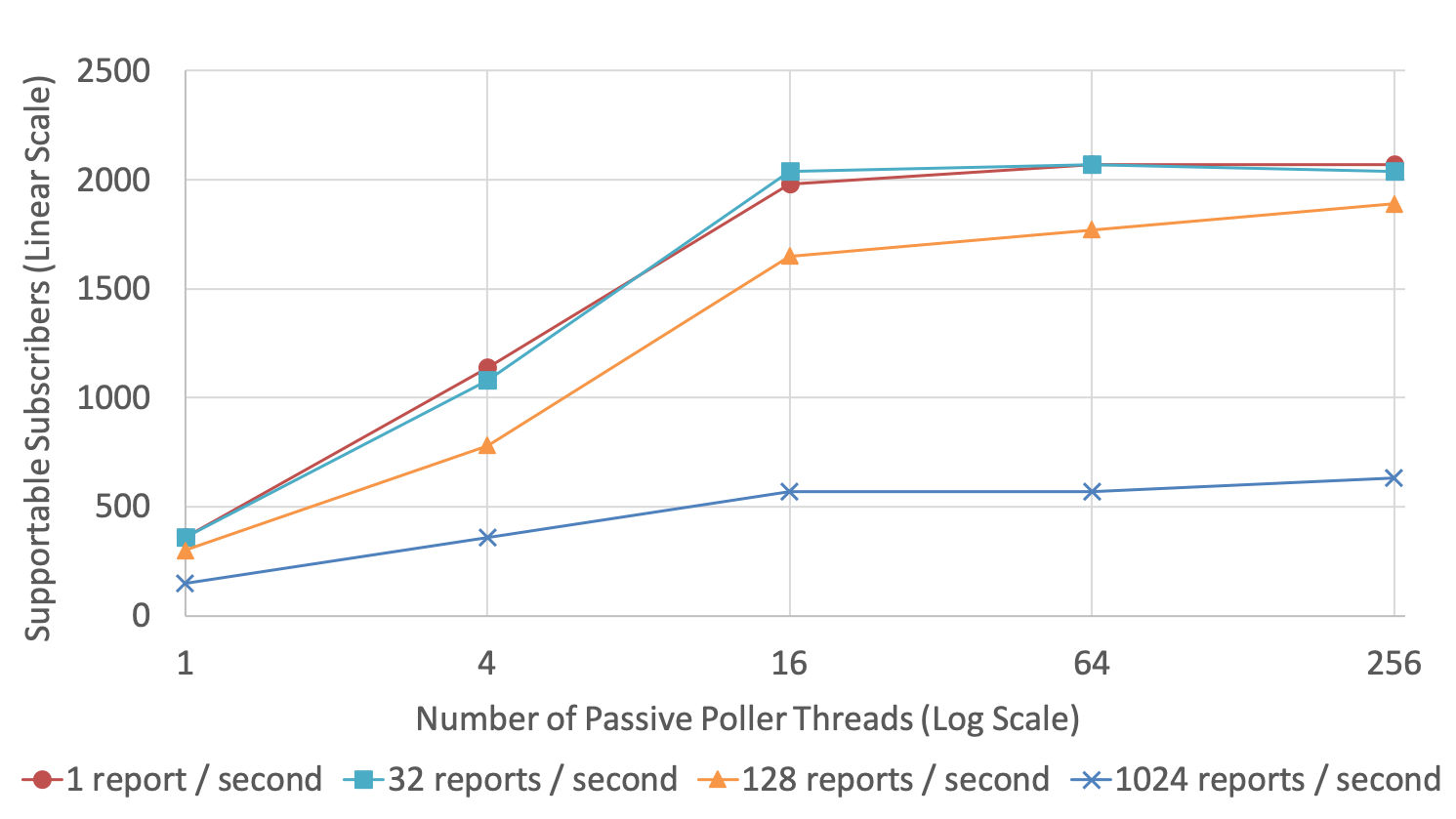}
    \caption{Supportable Subscriber gains as the number of polling threads is increased (Using Case 1)}
  \label{ThreadComparisonGraph}
\end{figure}

Increasing the number of pollers allowed multiple queries to run concurrently, thus resulting in better performance, using however more of the system's resources than the BAD channel query that runs repeatedly. The performance gains will be limited by the additional query compilation cost. To show this limitation, we conducted an experiment (using the Case 1 scenario) where we varied the number of pollers and measured the performance in terms of supportable subscribers. The results are shown in Figure~\ref{ThreadComparisonGraph}, for different report rates. 
As can be seen, the passive mode can only be improved to a limited extent by adding more pollers; for all examined report rates, the performance soon flattens after 16 pollers. Furthermore, the maximum number of supportable subscribers achieved decreases as the report rate increases. For the lower report rates (1 and 32 reports/sec) the main bottleneck is the compilation cost of the polling queries. For the higher report rates (128 and 1024 reports/sec) the performance is affected by the compilation cost as well as the increased computational cost introduced by having more reports. 

It should be noted that the very high report rates in our experiments were used so as to identify the limits of the BAD system, rather than representing a realistic scenario. High rates with thousands or even hundreds of emergencies per second would create a practically unusable amount of data for each subscriber. A subscriber would be expected to read hundreds or thousands of results during one execution (every 10 seconds). In more `realistic' scenarios, where subscribers might each get one or a few notifications during an execution, the BAD system performs orders of magnitude better than the passive mode.

\subsection{Diving into the BAD Performance Details} 
\label{whereTimeGoes}
\begin{figure*}
    \centering
    \begin{subfigure}[b]{0.32\textwidth}
        \includegraphics[width=\textwidth]{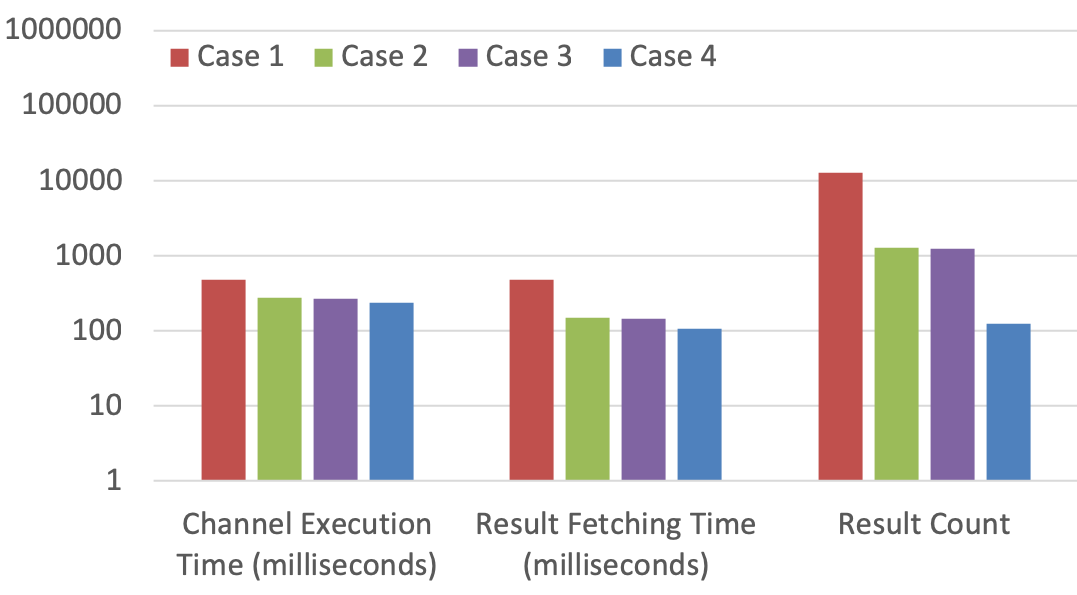}
        \caption{900 users with 4 reports / second}
        \label{fig:gull}
    \end{subfigure}
    ~ 
    \begin{subfigure}[b]{0.32\textwidth}
        \includegraphics[width=\textwidth]{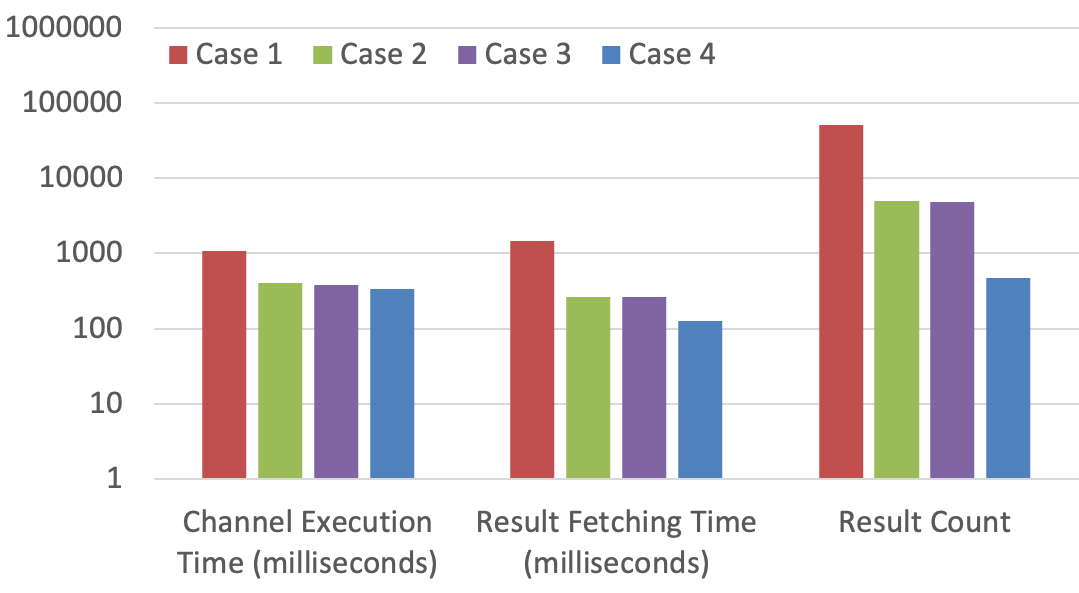}
        \caption{900 users with 16 reports / second}
        \label{fig:tiger}
    \end{subfigure}
    ~ 
    \begin{subfigure}[b]{0.32\textwidth}
        \includegraphics[width=\textwidth]{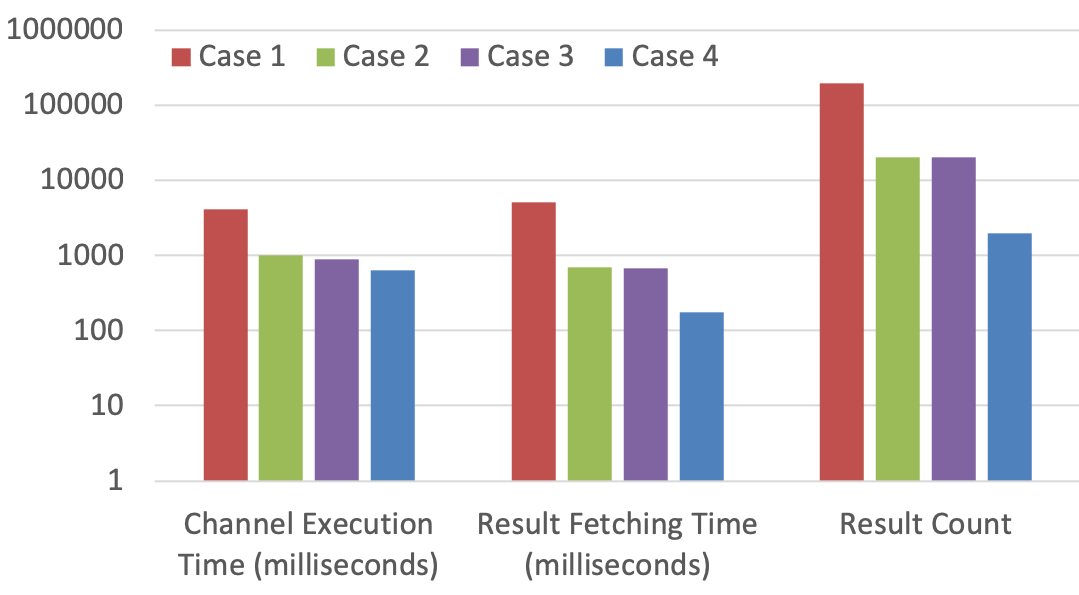}
        \caption{900 users with 64 reports / second}
        \label{fig:mouse}
    \end{subfigure}
    \caption{Channel statistics with different rates of reports in all four cases}
    \label{fig:stats}
\end{figure*}

In this section, we look a bit more deeply at the performance details of the channel execution in different cases to investigate the factors impacting the performance of the BAD system.
To do so, we fixed the number of subscribers at 900, the window size was again 10 seconds and we used three report rates, namely 4, 16 and 64 reports/sec. We report the channel execution time, the broker result fetching time, and the result size (number of records) averaged over 30 executions. The results are shown in Figure~\ref{fig:stats}.

The channel execution time, result fetching time, and result count all increase as the report rate increases. This is because there are more reports contributing to the channel query evaluation. In all three settings, the channel execution time of Case 1 is higher than Cases 2 and 3, which are slightly higher than Case 4. The reason is that in Case 1, there is a higher probability that a user's location intersects with an emergency report. This in turn leads to a higher result size and thus a higher result fetching time.
In Case 2, we moved 90\% of the reports to another city, and in Case 3, we moved 90\% of the users to another city. Both cases reduced the size of the data participating in result generation by the same factor of 0.9. Thus, we see similar execution times in both cases, and their result size and result fetching times are roughly the same as well.
In Cases 2 and 3, as we moved most of the reports or the users to a different city, both the evaluation time and result size become smaller than Case 1. In Case 4, there is only a small fraction of users and emergency reports that could intersect, so the channel execution time, result fetching time and result size are the smallest.

\subsection{BAD vs. Postgres}
In discussing BAD with various audiences, we have found ``what about triggers?" to be a frequently-asked question. As a result, to explore the challenges of handling the emergencies-near-me use case without a system like BAD, but with a single system -- i.e., without resorting to gluing multiple systems together -- we also attempted to achieve the same goals using a DBMS that supports triggers. 
We chose Postgres \cite{stonebraker1986design} for this exercise due to its popularity and long-standing support for triggers.
Since Postgres is a single node
DBMS\footnote{There is a distributed variant of Postgres -- from Greenplum -- that provided database triggers in an earlier version. However, triggers have been removed in the current version due to their unreliable behavior in a distributed setting \cite{greenplum}.}
we could only compare its performance against a single node deployment of BAD.
 
Recall that the emergencies-near-me scenario has two tables (datasets) being updated actively, \emph{UserLocations}
and \emph{Reports}.
Data changes in either table could generate new notifications for corresponding subscribers, e.g., a user walks into an emergency event or a new emergency event happens near a subscribed user. 
In order to capture both cases, we created one trigger on each of these Postgres tables. In response to an update/insert on \emph{UserLocations},
the corresponding trigger finds intersecting reports that happened in the past 10 seconds, joins them with the \textit{Shelters} table, and inserts the result (i.e., a new notification) into the \emph{Results} table. The trigger on \emph{Reports} works in a similar way for each new report insertion into that table. 

Special care is needed to avoid duplicate results. Consider the following scenario. Assume that the last known location of a user is reported at time $t_0$. If an emergency that happens at time $t_1$ (where $t_1 > t_0$) intersects with this user's location, the \textit{Reports} trigger will produce a new result. Soon after, at time $t_2$ (where $t_2 - t_1 < 10$) that user sends a new location update which also intersects with the same emergency. The \textit{UserLocations} trigger will produce a (duplicate) result with the same emergency and user id. Such duplication will increase the result size and lead to the users receiving redundant notifications. 
To avoid this we defined the \emph{Results} primary key as $<user\_name, report\_id>$ and set the triggers to upsert data into this table. 

In this experiment, we created two client programs that feed separately the \emph{UserLocations} and \emph{Reports} tables 
by issuing upsert and insert statements respectively. 
The client for \emph{UserLocations} upserts user locations every 10 seconds. The client for the \emph{Reports} table inserts new reports at a specified rate (as needed by the experiment). Unlike the BAD data feeds, both clients could be slowed down due to expensive trigger calls and the overall system load. 
We consider the trigger-based approach as being \textit{overloaded} if any of the clients fails to upsert/insert data at the specified rate.
For the experiment, we also set up an external broker program that pulls recent results from the system every 10 seconds. 

For comparison purposes, we also implemented a passive mode with one poller thread using Postgres. Similar to the trigger implementation, we also created the \emph{Reports} table and set up an external client program that sends new reports at a specified rate. Instead of relying on triggers to generate results and having a broker program to fetch them, we set up a poller program that polls the nearby emergencies and shelters on behalf of each user. This poller works in the same way as the passive mode of AsterixDB in Section~\ref{bad_expr}. Similarly to the trigger approach, both the \emph{Reports} client and the poller program could be slowed by the system load. We consider the one poller implementation as able to support a certain amount of users and rate of reports if both the \emph{Reports} client and the poller program can update/query data at the specified rate.

We deployed the single node BAD system with both CC and NC on the same node. We created the same datatypes, datasets, and channels from Section~\ref{bad_expr}. Note that while Postgres cannot be scaled to multiple nodes, the BAD system's performance can be improved by scaling in a larger cluster. For comparison purposes, we also added experimental results based on running BAD on 4 nodes (one CC and three NCs).

\begin{figure}[!ht]
  \centering
  \includegraphics[width=0.45\textwidth]{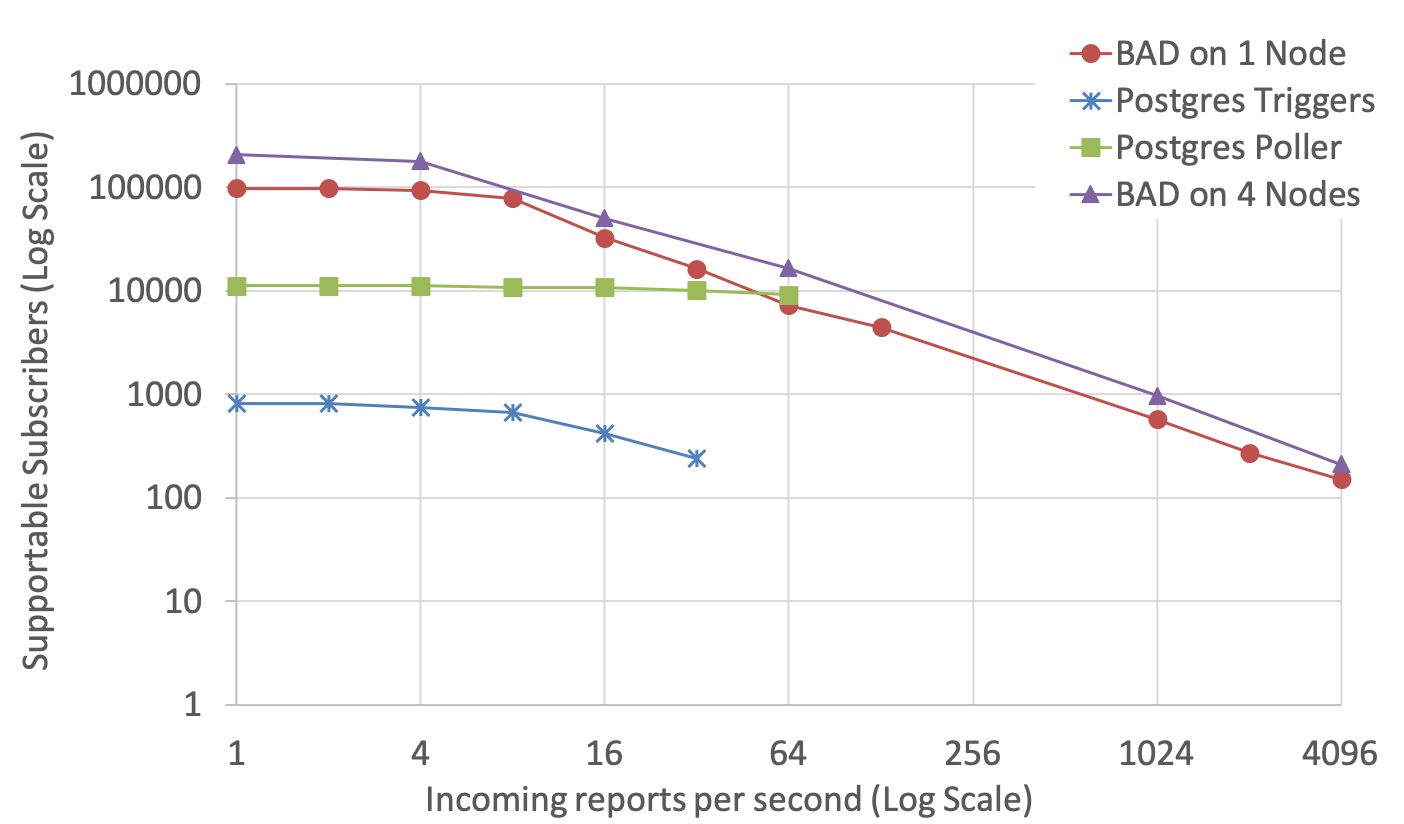}
    \caption{Emergencies-near-me on Postgres vs. BAD}
  \label{badpostgres}
\end{figure}

The experimental results of the comparison (in logarithmic scale) are shown in Figure~\ref{badpostgres}; the scenario used was Case 4. In the figure ``Postgres Triggers'' and ``Postgres Poller'' denote the triggers and single-poller implementations on Postgres respectively. Similarly, ``BAD on 1 node'' and ``BAD on 4 nodes'' correspond to the single- and four- node deployments of BAD. Both the ``Postgres Triggers'' and ``Postgres Poller'' start with relatively stable performance. ``Postgres Triggers'' starts declining as the rate of reports gets higher because more incoming reports make the computation in both triggers more expensive. 
As the report rate increases further (above 32 reports/sec for the ``Postgres Triggers'' and 64 reports/sec for the ``Postgres Poller''), the \emph{Reports} clients in the Postgres-based implementations 
fail to add new data at the specified rate. 

In contrast to Postgres, ``BAD on 1 node'' starts with a much higher number of supportable users. Benefiting from the data feeds and channel mechanisms, BAD is able to consume rapid incoming data and produce results, even when the report arrival rate is very high. 
Further improvements can be achieved by scaling. (See ``BAD on 4 nodes'' and remember the log scale.)

We finally note that in the ``Postgres Triggers'' case, we have to create triggers for each of the ``active'' tables that participates in the computation. This adds extra complexity for application development (managing duplicate results, etc.) However, in BAD, a single channel can access many datasets and compute complex results on their data. The BAD system not only provides better performance compared with the traditional trigger implementation, but it also reduces the effort involved in building BAD applications.

\subsection{Cluster Scaling Experiments}
\textbf{Speed-up:} In order to show how the channel execution time can scale with the cluster size, we performed additional experiments on a 20-node cluster with Dual-Core AMD Opteron Processor 2212 2.0GHz. Each machine had 8 GB of memory and dual 1 TB 7200 RPM SATA disks. Recall that a BAD cluster consists of one CC node and one or more NC nodes. The number of NC nodes determines the computational power of the cluster when executing jobs (including the channel execution). We conducted this experiment using Case 1, but extended the channel execution window and query (as well as the update rate of users' locations) to 20 seconds. We tested three different scenarios: 4140 users with 16 reports/sec, 2160 users with 32 reports/sec, and 1020 users with 64 reports/sec. The number of users setting was determined by finding the maximum number of supportable users on the cluster with two NC Nodes for the given report rate. 

\begin{figure}[!ht]
  \centering
  \includegraphics[width=0.45\textwidth]{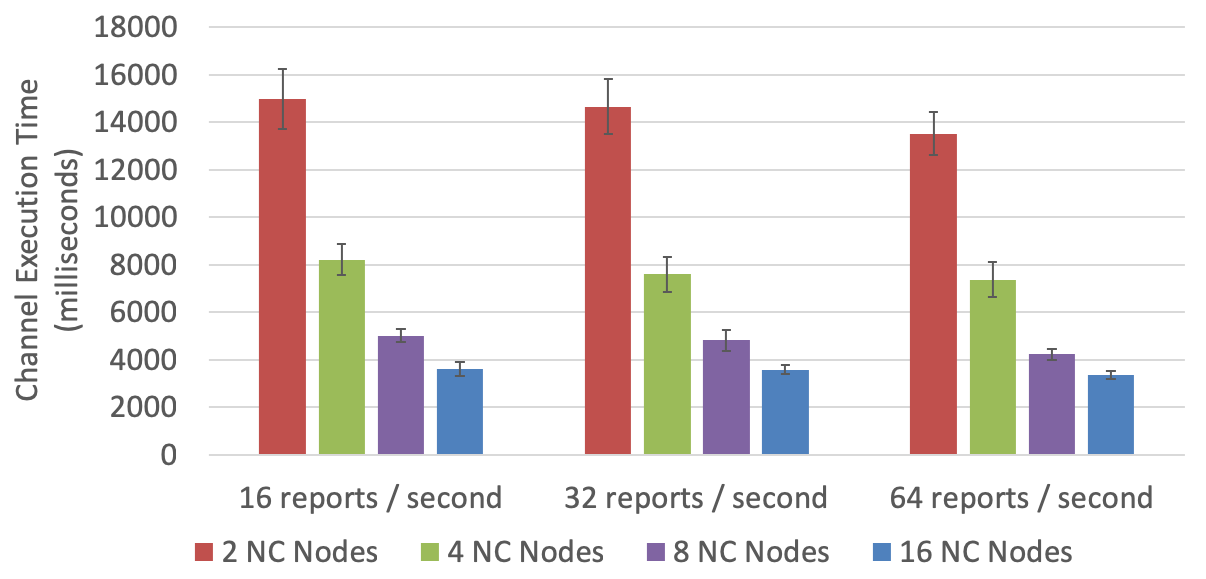}
    \caption{Channel speed-up for different cluster sizes}
  \label{cluster}
\end{figure}

We measured the channel execution time on 4 different scales (2 NC nodes, 4 NC nodes, 8 NC nodes, and 16 NC nodes). The other configurations remained the same as Section~\ref{bad_expr}. We show the speed-up performance against the number of NC nodes in Figure~\ref{cluster}. All results are reported with 95\% confidence. As expected, the channel execution time decreases as we introduce more nodes, showing good speed-up performance.

\begin{figure}[!ht]
  \centering
  \includegraphics[width=0.45\textwidth]{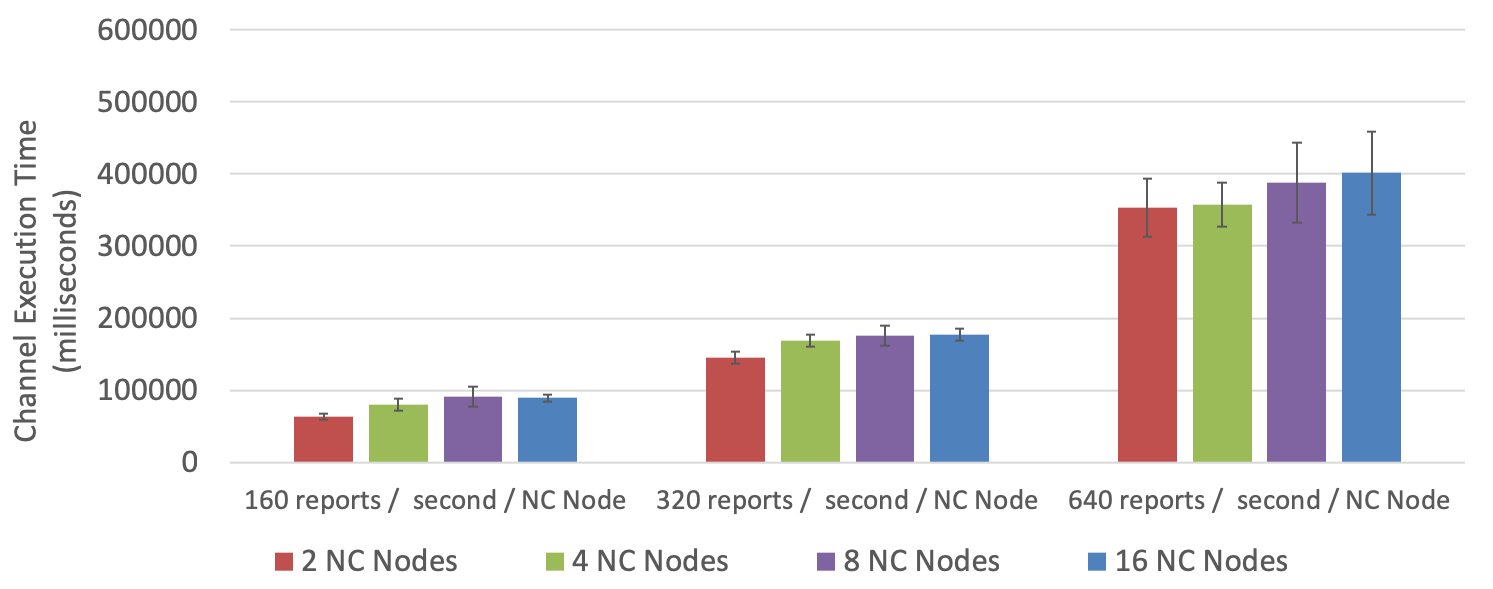}
    \caption{Channel scale-up for different cluster sizes}
  \label{cluster2}
\end{figure}

\textbf{Scale-up:} We also measured the scale-up performance by increasing the report rates in proportion to the cluster size using Case 1. We tested with 900 users and three different report rates per NC node (160 reports/sec/node, 320 reports/sec/node, and 640 reports/sec/node) on 4 different scales (2 NC nodes, 4 NC nodes, 8 NC nodes, and 16 NC nodes). In order to better highlight large scales of data and cluster utilization we used a 600 second channel execution window. The other configurations remained the same as in Section~\ref{bad_expr}. The results are shown in Figure~\ref{cluster2}. All results are reported with 95\% confidence. As we increased the cluster size and report rate together, the channel execution time maintained relative stability and only grew slightly due to the increased execution overhead of a larger cluster. This shows that the BAD system can scale well for use cases with larger workloads.

\section{Conclusions and Future work} \label{conclusion}
We have introduced a new paradigm for big data, namely \textit{Big Active Data}, that merges Big Data Management with active capabilities. We have implemented a BAD system prototype using a modern Big Data Platform (AsterixDB), and we showed how it can outperform passive Big Data by an order (or two) of magnitude in many practical scenarios. BAD can consider data in context and enrich results in ways unavailable to other active platforms, in addition to allowing for retrospective Big Data analytics.
Our code ($>$ 20,000 LOC) is available as an open-source Apache project \cite{bad:code}.

From this point, BAD can be improved in a myriad of ways. For example, this paper only scratched the surface of the research for Broker to User communication and scalability. In addition, we currently treat channels as completely isolated jobs. We can instead tackle the task of scaling multiple channels together by recognizing common work (e.g., detecting recent emergencies) and sharing this work between channels at the runtime level.
While we focus here on repetitive channels, they are limited by the periodicity that they need to execute at. As future work we plan to create \emph{continuous channels}, channels that will execute based on data changes as they happen rather than on fixed intervals.

We also believe that BAD is ready for a rich performance benchmark. This paper has focused on the big picture and initial results of BAD, and therefore is not a comprehensive look at all performance optimization possibilities. 
For example, based on our experimental results, we have seen that the overhead of staging the results on the data cluster can be a limiting factor for performance in some cases. 
We are currently exploring a push-based channel model where results are more eagerly sent directly to brokers (rather than just notifications of results).
We are also working on a comparison of the BAD approach with a glue-based approach.
Experiments on a much larger cluster with higher-scale workloads would also be an interesting future undertaking.

\begin{acknowledgements}
This research was partially supported by NSF grants IIS-1447826, IIS-1447720,  IIS-1838222, IIS-1838248, CNS-1924694 and CNS-1925610.
\end{acknowledgements}

\bibliographystyle{abbrv}
\bibliography{citations}   

\end{document}